\documentclass[structabstract]{aa}  
\usepackage{graphicx}
\usepackage{txfonts}
\usepackage{natbib}
\usepackage{lscape}
\usepackage{afterpage}
\usepackage{dblfloatfix}

\PassOptionsToPackage{linktocpage}{hyperref}
\usepackage{color}

\begin{document}

   \title{Gas-phase CO depletion and N$_\mathsf{2}$H$^+$ abundances in starless cores\thanks{This work is partially based on observations by the Herschel Space Observatory. {\it Herschel} is an ESA space observatory with science instruments provided by European-led Principal Investigator consortia and with important participation from NASA.}}

   \author{N. Lippok\inst{1}\and
          R. Launhardt\inst{1}\and
          D. Semenov\inst{1}\and
          A. M. Stutz\inst{1}\and
          Z. Balog\inst{1}\and
          Th. Henning\inst{1}\and
          O. Krause\inst{1}\and
          H. Linz\inst{1}\and
          M. Nielbock\inst{1}\and
          Ya. N. Pavlyuchenkov\inst{2}\and
          M. Schmalzl\inst{3}\and
          A. Schmiedeke\inst{4}\and
          J. H. Bieging\inst{5}
          }

   \institute{Max-Planck-Institut f\"ur Astronomie (MPIA), K\"onigstuhl 17, D-69117 Heidelberg, Germany\\ \email{lippok@mpia.de}\label{inst1}
   \and Institute of Astronomy, Russian Academy of Sciences, Pyatnitskaya str. 48, Moscow 119017, Russia\label{inst2}
   \and Leiden Observatory, Leiden University, PO Box 9513, 2300 RA, Leiden, The Netherlands\label{inst3}
   \and Universit\"at zu K\"oln, Z\"ulpicher Str. 77, D-50937 K\"oln, Germany
   \and Department of Astronomy and Steward Observatory, University of Arizona, Tucson, AZ 85721, USA\label{inst5}}

   \date{Received June 24, 2013; accepted September 13, 2013}

 
  \abstract
  {In the dense and cold interiors of starless molecular cloud cores, a number
    of chemical processes allow for the formation of complex molecules and
    the deposition of ice layers on dust grains. Dust density and temperature
    maps of starless cores derived from Herschel continuum observations
    constrain the physical structure of the cloud cores better than ever before. We use these to model the temporal chemical evolution of starless cores.}
   {We derive molecular abundance profiles for a sample of starless cores. We then analyze these using chemical modeling based on dust temperature and hydrogen
     density maps derived from Herschel continuum observations.}
   {We observed the $^{12}$CO\,(2-1), $^{13}$CO\,(2-1), C$^{18}$O\,(2-1) and N$_2$H$^+$\,(1-0) transitions towards seven isolated, nearby low-mass
     starless molecular cloud cores. Using far infrared~(FIR) and submm dust emission maps
     from the Herschel key program Earliest Phases of Star formation~(EPoS)
     and by applying a ray-tracing technique, we derived the physical
     structure~(density, dust temperature) of these cores. This physical
     structure is then used to derive molecular abundance profiles from the
     molecular line data. In addition, we applied time-dependent chemical and
     line-radiative transfer modeling and compared the modeled to the observed
     molecular emission profiles.}
   {CO is frozen onto the grains in the center of all cores in our sample. The level of CO depletion increases with hydrogen density and ranges from 46\% up to more than 95\% in the core centers of the three cores with the highest hydrogen density. The average hydrogen density at which 50\% of CO is frozen onto the grains is $1.1\pm 0.4\times 10^5$\, cm$^{-3}$. At about this density, the cores typically have the highest relative abundance of N$_2$H$^+$. The cores with higher central densities show depletion of N$_2$H$^+$ at levels of 13\% to 55\%. The chemical ages for the individual species are on average $(2\pm 1)\times 10^5$~yr for $^{13}$CO, $(6\pm 3)\times 10^4$~yr for C$^{18}$O, and  $(9\pm 2)\times 10^4$~yr for $\mathrm{N_2H^+}$. Chemical modeling indirectly suggests that the gas and dust temperatures decouple in the envelopes and that the dust grains are not yet significantly coagulated.}
  {We observationally confirm chemical models of CO-freezeout and nitrogen chemistry. We find clear correlations between the hydrogen density and CO depletion and the emergence of $\mathrm{N_2H^+}$. The chemical ages indicate a core lifetime of less than 1\,Myr.}


   \maketitle
%


\section{Introduction}
\renewcommand\floatpagefraction{0.5}
\renewcommand\topfraction{0.5}

Stars form in cold cloud cores. In the prestellar
phase, the centers of the cloud cores cool down to temperatures below 10\,K and
the hydrogen densities typically exceed $10^5$\,cm$^{-3}$. In these conditions
complex molecules are synthesized~\citep{bacmann2012}. At the same time, in
the densest regions of the cores, many molecular species stick onto the dust grains~\citep[e.g.,][]{bergin2007}. The evolution of the chemical composition influences the cooling rate of the clouds and changes the degree of ionization,
both important properties that influence the star formation process.

The gas in the local interstellar medium (ISM) consists of 98\% molecular hydrogen and helium. These species, however, do not emit at cold temperatures. Therefore, CO has become a popular tracer of the ISM. It is the second most abundant molecule and readily excited already at low temperatures. However, in the conditions in prestellar cores, CO freezes out onto the dust grains. This alters the chemical composition of the cloud cores and the dust properties. Nitrogen-bearing species like N$_2$H$^+$, which are otherwise destroyed by CO molecules, can now accumulate in the gas phase and trace these denser regions. N$_2$H$^+$ molecules form and deplete on longer time scales than does CO. Under reasonable assumptions of the initial abundances, physical conditions, and reaction rates, modeling both species together can therefore constrain the chemical age of the cores.

While early studies had already detected CO mantles on dust grains from
absorption features in the mid-infrared \citep[e.g.,][]{tielens1991},
direct observational proof of molecular depletion from the gas phase in the centers
of dense cores came only later. It requires the accurate determination of molecular abundances in the gas phase and of the total gas mass. One of the most robust ways to determine the total mass in dense clouds is to infer it via a presumably known gas-to-dust mass ratio from (column-) densities of the dust. They in turn can be derived from the black-body radiation of the dust grains~\citep[e.g.,][]{launhardt2013} or the visual to mid-infrared extinction caused by the dust~\citep[e.g.,][]{witt1990,alves2001}.

Several studies have been performed that quantify the CO depletion in starless cores on the basis of hydrogen masses deduced from dust measurements~\citep[e.g.,][]{willacy1998,caselli1999,bacmann2002,bergin2002,tafalla2004,pagani2005,stutz2009,ford2011}. All of them find strong indications of a central freezeout of CO. So far, however, the observational studies of molecular depletion have only had weak constraints on the temperature distribution within molecular clouds or none at all. Knowledge of the dust temperature is, however, essential for both derivation of column densities from the dust emission and robust
chemical modeling~\citep{pavlyuchenkov2007}. Uncertainties in the temperature also introduce large errors into the hydrogen densities derived from (sub-)mm dust emission.

Isolated Bok globules, which often contain only a single dense core
and are devoid of larger envelopes, are the best suited objects
for detailed studies of the physical and chemical structure of starless cores. We therefore observed seven globules containing a starless core as part of the \textit{Herschel} Guaranteed Time Key project EPoS. The \textit{Herschel} space observatory \citep{pilbratt2010} and its
sensitive PACS \citep{poglitsch2010} and SPIRE
\citep{griffin2010} bolometer arrays have made it possible
to sample the peak of the thermal spectral energy distributions~(SED) of cold molecular clouds with high
sensitivity and spatial resolution for the first time. In \citet{stutz2010,launhardt2013} we presented dust temperature and hydrogen density maps of the starless cores in the globules derived by modified black-body
fits to the continuum emission. We also developed a ray-tracing
method to restore the full volume density and dust temperature structure of these cores. The results are presented in \citet{nielbock2012} for B\,68, for CB\,17 in \citet{schmalzl2013}, and in this paper for the remaining five globules.

We also present maps of the $^{12}$CO~(J=2-1), $^{13}$CO~(J=2-1), C$^{18}$O~(J=2-1), and N$_2$H$^+$~(J=1-0) transitions of the same seven globules. The maps of $^{12}$CO~(J=2-1) and $^{13}$CO~(J=2-1) of the globules CB\,17, CB\,26, and CB\,244 were previously presented in \citet{stutz2009}. Based on dust temperature and hydrogen density maps derived from the \textit{Herschel} observations, and using time-dependent chemical modeling, we derive the molecular abundances. From the modeling results, we also constrain the chemical age of the cores and infer information on the dust grains in the globules. 

The paper is structured as follows. In Sect.~2 we describe the observations and data reduction. In Sect.~3 we describe our modeling approaches. In Sect.~4 we present and discuss the results. We summarize our findings in Sect.~5. Maps of the observations and the derived dust density and hydrogen density maps are shown in the online appendix. We also present a simple LTE analysis of the molecular column densities in Appendix~\ref{sec:LTE}.

\begin{table*}
\caption{\label{t1}\footnotesize{Source list}}
\centering
\begin{tabular}{llclcc}
\hline\hline
Source&Other&R.A., Dec. (J2000)\tablefootmark{a}&Region&Dist.&Ref.\\
&names&[h:m:s, $^\circ$:$\arcmin$:$\arcsec$]&&[pc]\\
\hline
CB 4  &  ...      &  00:39:05.2, +52:51:47  &  Cas A, Gould Belt (GB)   &  $350\pm 150$  & 1\\
CB 17\tablefootmark{b} & L 1389    &  04:04:37.1, +56:56:02  &  Perseus, GB &  $250\pm 50$   & 2,
1\\
CB 26\tablefootmark{c} & L 1439    &  05:00:14.5, +52:05:59  &  Taurus-Auriga &$140\pm 20$   & 2, 3,4\\
CB 27 & L 1512    &  05:04:08.1, +32:43:30  &  Taurus-Auriga &$140\pm 20$   & 3,4,5\\
B68   & L 57, CB 82 &17:22:38.3, -23:49:51  &  Ophiuchus, Pipe nebula & $135\pm
15 $ & 6, 7, 8 \\
CB 130\tablefootmark{d} & L 507    &  18:16:14.3, -02:32:41  &  Aquila rift, GB &$250\pm 50$ & 9,
10 \\
CB 244\tablefootmark{e} & L 1262   &  23:25:26.8, +74:18:22  &  Cepheus flare, GB &$200\pm 30$ &
2, 4, 11 \\
\hline \hline
\end{tabular}
\tablefoot{
\tablefoottext{a}{Positions of the center of the starless cores, defined as the column density peaks found in \citet{launhardt2013}.} 
\tablefoottext{b}{CB 17: additional low-luminosity Class I YSO $25\arcsec$ from the starless core.}
\tablefoottext{c}{CB 26: additional Class I YSO $3.6\arcmin$ southwest of the
  starless core.}
\tablefoottext{d}{CB 130: additional Class 0 core $\sim 30\arcsec$ east and
  Class I YSO $\sim 45\arcsec$ east of starless core.}
\tablefoottext{e}{CB 244: additional Class 0 source $\sim 90\arcsec$ east of the starless core.}
}
\tablebib{(1)~\citet{perrot2003}; (2)~\citet{launhardt2010}; (3)~\citet{loinard2011}; (4)~\citet{stutz2009};
  (5)~\citet{launhardt2013}; (6)~\citet{deGeus1989};
  (7)~\citet{lombardi2006}; (8)~\citet{alves2007}; (9)~\citet{launhardt1997a}; (10)~\citet{straizys2003}; (11)~\citet{kun1998}.
}
\end{table*}

\section{Observations}

\begin{table*}
\caption{\label{tab_obs}\footnotesize{List of observations}}
\centering
\begin{tabular}{llccccccccc}
  \hline\hline
  Source & Line & Freq. & Tel. & Date & HPBW & $\Delta \mathrm{v}$ & $\eta_\mathrm{mb}$
  & Map size & Ref. \\
  & & [GHz] & & [mo/yr] & [arcsec] & [m s$^{-1}$]\\
  \hline
  CB 4 & $^{12}$CO\,(2--1)          &    230.537990 & HHT  & 10/09 & 32.2 &    300 &
  0.85 &10\arcmin $\times$ 10\arcmin \\
  &$^{13}$CO\,(2--1)          &    220.398686 & HHT  & 10/09 & 33.7 &    300 &
  0.85 &10\arcmin $\times$ 10\arcmin \\
  &C$^{18}$O\,(2--1)          &    219.560319 & 30m  & 12/10 & 11.3 &    53
  & 0.52 & 3\arcmin $\times$ 3\arcmin \\
  &N$_2$H$^+$\,(1--0)        &  93.1737       & 30m  & 12/10 & 26.6 &   63
  & 0.75 & 3\arcmin $\times$ 3\arcmin\\
  \hline
  CB 17 & $^{12}$CO\,(2--1)          &    230.537990 & HHT  & 04/08 & 32.2 &    300 &
  0.85 &10\arcmin $\times$ 10\arcmin \\
  & $^{13}$CO\,(2--1)          &    220.398686 & HHT  & 04/08 & 33.7 &    300 &
  0.85 &10\arcmin $\times$ 10\arcmin \\
  & C$^{18}$O\,(2--1)          &    219.560319 & 30m  & 10/96 & 10.9 &
  107 & 0.48 & 35\arcsec / 7\arcsec \\
  & N$_2$H$^+$\,(1--0)        &  93.1737       & 30m  & 10/96 & 26.6 & 63 &
  0.95 & $140\arcsec \times 140 \arcsec$\\
  \hline
  CB 26 & $^{12}$CO\,(2--1)          &    230.537990 & HHT  & 03/09 & 32.2 &    300 &
  0.85 &10\arcmin $\times$ 10\arcmin & 1\\
  & $^{13}$CO\,(2--1)          &    220.398686 & HHT  & 03/09 & 33.7 &    300 &
  0.85 &10\arcmin $\times$ 10\arcmin & 1\\
  & C$^{18}$O\,(2--1)          &    219.560319 & 30m  & 12/10 & 11.3 &    53
  & 0.52 & 6\arcmin $\times$ 5\arcmin \\
  & N$_2$H$^+$\,(1--0)        &  93.1737       & 30m  & 12/10 & 26.6 &   63
  & 0.75 & 6\arcmin $\times$ 5\arcmin\\
  \hline
  CB 27 & $^{12}$CO\,(2--1)          &    230.537990 & HHT  & 01/07 & 32.2 &    300 &
  0.85 &10\arcmin $\times$ 10\arcmin & 1\\
  & $^{13}$CO\,(2--1)          &    220.398686 & HHT & 01/07 & 33.7 &    300 &
  0.85 &10\arcmin $\times$ 10\arcmin & 1\\
  & C$^{18}$O\,(2--1)          &    219.560319 & 30m  & 12/10 & 11.3 &    53
  & 0.52 &3\arcmin $\times$ 3\arcmin\\
  & N$_2$H$^+$\,(1--0)        &  93.1737       & 30m  & 12/10 & 26.6 &   63
  & 0.75 & 3\arcmin $\times$ 3\arcmin\\
  \hline
  B 68 & $^{12}$CO\,(2--1)          &    230.537990 & HHT  & 03/08 & 32.2 &    300 &
  0.85 &10\arcmin $\times$ 10\arcmin \\
  & $^{13}$CO\,(2--1)          &    220.398686 & HHT  & 03/08 & 33.7 &    300 &
  0.85 &10\arcmin $\times$ 10\arcmin \\
  & C$^{18}$O\,(1--0)         & 109.782182 & 30m & 04/00 & 23.6 & 30 & 0.91
  & 4\arcmin $\times$ 4\arcmin / 12\arcsec\tablefootmark{(b)}\\
  & N$_2$H$^+$\,(1--0)   &  93.1737       & 30m  & 04/01 & 26.6 &  63
  & 0.75 &3\arcmin $\times$ 3\arcmin / 12-24\arcsec\tablefootmark{(b)}  \\
  \hline
  CB 130 & $^{12}$CO\,(2--1)          &    230.537990 & 30m  & 07/11 & 11.3 &    51 &
  0.92 &4\arcmin $\times$ 4\arcmin \\
  & $^{13}$CO\,(2--1)          &    220.398686 & 30m & 07/11 & 11.8 &    53 &
  0.92 &4\arcmin $\times$ 3\arcmin \\
  & C$^{18}$O\,(2--1)          &    219.560319 & 30m  & 01/11 & 11.3 &    53
  & 0.52 & 5\arcmin $\times$ 5\arcmin\\
  & N$_2$H$^+$\,(1--0)        &  93.1737       & 30m  & 01/11 & 26.6 &   63
  & 0.75 &5\arcmin $\times$ 5\arcmin\\
  \hline
  CB 244 & $^{12}$CO\,(2--1)          &    230.537990 & HHT  & 04/08 & 32.2 &    300 &
  0.85 &10\arcmin $\times$ 10\arcmin & 1\\
  & $^{13}$CO\,(2--1)          &    220.398686 & HHT  & 04/08 & 33.7 &    300 &
  0.85 &10\arcmin $\times$ 10\arcmin & 1\\
  & C$^{18}$O\,(2--1)          &    219.560319 & 30m  & 12/10 & 11.3 &    53
  & 0.52 &5\arcmin $\times$ 5\arcmin\\
  & N$_2$H$^+$\,(1--0)        &  93.1737       & 30m  & 12/10 & 26.6 &   63
  & 0.75 & 5\arcmin $\times$ 5\arcmin\\
  \hline \hline
\end{tabular}
\tablefoot{
  \tablefoottext{a}{At the center of the starless core.}
  \tablefoottext{b}{Data first presented in \cite{bergin2002}. These maps were obtained from tracked observations with the
   listed pointing separations.}
  \tablebib{(1)~\citet{stutz2009}}
}
\end{table*}

\subsection{Source selection}

Within the EPoS project, 12 nearby Bok globules have been observed with the \textit{Herschel}
bolometers. The sample of globules has been selected based on results from
previous studies \citep[see][and references therein]{launhardt2013} and was
known to contain low-mass pre- and protostellar cores. Of particular importance for the selection was the criterion that the globules be very isolated and
located outside the galactic plane. Their mean galactic latitude is $7.3\pm 3$~degrees,
with the closest one still 3.5~degrees away from the galactic plane, such that
background levels and confusion from the galactic plane are minimized. Out of the 12 globules,
seven globules contain starless cores. These globules are studied in this paper and are listed in Table~\ref{t1}. Three of the globules contain only a single starless core (CB~4, CB~27, and
B~68), one contains an additional Class 0 core (CB~244), and two contain an additional Class I
young stellar object (CB~17 and CB~26). CB~130 contains two additional cores
(Class 0 and I).

\subsection{Continuum observations}

All globules were observed in the \textit{Herschel} PACS bands at 100 and
160\,$\mu$m and the \textit{Herschel} SPIRE bands at 250\,$\mu$m, 350\,$\mu$m, and 500\,$\mu$m. These
observations were complemented with ground-based (sub-)~mm observations
ranging from 450\,$\mu$m to 1.2\,mm. The respective observations and
data reduction are described in detail in \citet{nielbock2012} and \citet{launhardt2013},
and references therein. The dust temperature and
density maps presented in Sect.~\ref{sec:dustresults} were derived from this set of observations.

\subsection{Molecular line observations}

On-the-fly maps of the C$^{18}$O\,(J=2-1) line at 219.560 GHz and the
N$_2$H$^+$\,(J=1-0) line complex around 93.174 GHz were taken at the IRAM 30m
telescope between November~2010 and January~2011 of all starless cores except B~68. For this globule, we used data that were obtained before and
published in \citet{bergin2002}. The observing parameters are compiled in
Table~\ref{tab_obs}. Map sizes are different for different sources and range from
$3\arcmin \times 3\arcmin$ to $6\arcmin \times 5\arcmin$. The HPBWs were
$12\arcsec$ for C$^{18}$O and $28\arcsec$ for N$_2$H$^+$. The channel widths
were 40\,kHz for C$^{18}$O and 20\,kHz for N$_2$H$^+$, corresponding to 0.05
and 0.06\,km/s. Both molecular transitions were observed
simultaneously using the EMIR receivers~\citep{carter2012}. Calibration was done using the
standard chopper-wheel method every 15 minutes. The weather conditions were
mixed during the course of the observations leading to mean system temperatures of 700\,K for C$^{18}$O and
180\,K for N$_2$H$^+$. During the observations of CB\,130, the weather was better,
resulting in lower system temperatures of 300\,K and 100\,K.

To minimize scanning artifacts, the on-the-fly maps were taken with two orthogonal
scan directions with two coverages each and a scanning speed of
$6\arcsec/\mathrm{s}$. A reference position was observed approximately every
1.5 minutes. These positions were derived from IRIS 100\,$\mu$m all-sky maps
\citep{miville2005}. They lie within extended regions of the lowest 100 $\mu$m continuum
emission, preferably within $10\arcmin$ from the source, and were expected to have no CO emission.

Molecular line maps of the $^{12}$CO and $^{13}$CO species at the $J=2-1$
transition~(230.538\,GHz and 220.399\,GHz) were obtained with the \textit{Heinrich
  Hertz Submilimeter Telescope} (HHT) on Mt.~Graham,~Arizona,~USA. The channel width was
$\sim0.34$~km/s and the angular resolution of the telescope was
32$''$~(\textit{FWHM}). For more details on the observations and data
reduction, we refer the reader to \citet{stutz2009}.

Table~\ref{tab_obs} lists all molecular line observations used for this study. All maps were reprojected to the coordinate
system of the corresponding dust temperature and density maps. They have also
been smoothed to the same Gaussian beam size of 36.4$\arcsec$ (resolution of the SPIRE~500\,$\mu$m maps and the dust temperature and density maps) and been regridded to
the same pixel scale of $10''\times 10''$.

\section{Modeling}
\subsection{Dust temperature and density maps from ray-tracing models}
\label{sec_rt_ana}

The goal of the ray-tracing fitting of the continuum observations of the
starless cores is to go beyond the line-of-sight (LoS)  optical depth-averaged analysis presented in \citet{launhardt2013} by also modeling the dust
temperature variation along the LoS and deriving volume density distributions.
The ray-tracing algorithm developed to derive the dust temperature and density structure of the
cores from the \textit{Herschel} and complementary ground-based continuum maps is
described in \citet{nielbock2012}. Here, we only briefly summarize the
hydrogen density and temperature profiles used by the algorithm and mention the formulae to illustrate the meaning
of the profile parameters.

For all modeling steps we use the dust opacities $\kappa_\mathrm{d}(\nu)$ tabulated in
\citet{oh94}\footnote{ftp://cdsarc.u-strasbg.fr/pub/cats/J/A+A/291/943} for
mildly coagulated composite dust grains with thin ice mantles ($10^5$~yrs coagulation time at gas density
$10^5$\,cm$^{-3}$)\footnote{This is not identical to the often-used OH5 opacities, which were calculated for a gas density of $10^6$\,cm$^{-3}$}. The density profiles of the cores have
been modeled assuming a ``Plummer''-like shape \citep{plummer1911},
which characterizes the radial density distribution of prestellar cores on the
verge of gravitational collapse \citep{whitworth2001}:
\begin{equation}
n_\mathrm{H}(r) = \frac{\Delta
  n}{\left(1+\left(\frac{r}{r_0}\right)^2\right)^{\eta/2}}+n_\mathrm{out} \;\;
\mathrm{if} \; r\le r_\mathrm{out} 
\label{eq_plummer}
\end{equation}
where $n_\mathrm H=2\times n(\mathrm H_2)+n(\mathrm H)$ is the total number density of hydrogen nuclei. The radius $r_\mathrm{out}$ sets the outer boundary of the
model cloud. The density beyond this radius is set to zero. We add a constant term
to the profile to account for the fact that the outer density and column density profiles in most globules do not drop off like a power law, but turn over into a low-density envelope.
This profile (i)~accounts for an inner flat density core
inside $r_0$ with a peak density $n_0=n_\mathrm{out}+\Delta n$~($\Delta n\gg n_\mathrm{out})$, (ii)~approaches modified power-law
behavior with an exponent $\eta$ at $r\gg r_0$, (iii)~turns over into a
flat-density halo outside $r_1$, where

\begin{equation}
\label{eqr1}
r_1=r_0\sqrt{\left(\frac{\Delta n}{n_\mathrm{out}}\right)^{2/\eta}-1}\,,
\end{equation}
and (iv)~is cut off at $r_\mathrm{out}$. The tenuous envelope
is actually neither azimuthally symmetric nor fully spatially recovered by our
observations. Therefore, its real size remains unconstrained, and
$r_\mathrm{out}$ cannot be considered a reliable source property.

For the temperature profile of an externally heated cloud, we adopt the
following empirical prescription. It resembles the radiation transfer
equation in coupling the local temperature to the effective optical depth
toward the outer ``rim'', where the interstellar radiation field~(ISRF)
affects

\begin{equation}
T(r)=T_\mathrm{out}-\Delta T\left(1-\mathrm{e}^{-\tau(r)}\right)
\label{eq:T}
\end{equation}
with $\Delta T = T_\mathrm{out}-T_\mathrm{in}$~and the frequency-averaged effective optical depth
\begin{equation}
\tau(r)=\tau_0 \frac{\int_{r}^{r_\mathrm{out}} n_\mathrm{H}(x)\,\mathrm{d}x}{\int_{0}^{r_\mathrm{out}}n_\mathrm H(x)\,\mathrm d x}.
\label{eq_tau}
\end{equation}
Here, $\tau_0$ is an empirical (i.e.,~free) scaling parameter that accounts
for the a priori unknown mean dust opacity and the spectral energy distribution of the UV radiation of the ISRF, and $T_\mathrm{in}$ is the minimum, inner temperature and is determined by IR and cosmic ray heating. 

The results of this modeling approach are presented in Sect.~\ref{sec:dustresults}. The uncertainties of the method have been discussed thoroughly in \citet{nielbock2012} and \citet{launhardt2013}. They come to the conclusion that the largest uncertainty~(up to a factor 3) in the derived density is introduced by the uncertainty in the dust opacities. Additional uncertainties are introduced by the assumed distances, the gas-to-dust ratio, the symmetry assumption (w.r.t. the plane-of-sky) for the LoS, and the limited angular resolution. In particular, in the center of the cores, the density can be systematically underestimated due to beam smoothing. We therefore estimate the overall fitting uncertainty of the derived density maps to be $\pm 25$\%, plus a scaling uncertainty by up to a factor 3 due to the not well-constrained dust opacity model. The accuracy of the dust temperature is not as strongly affected by these effects and the modeled dust temperature has an estimated uncertainty of $^{+2}_{-1}$~K.

\subsection{Chemical modeling}
\label{sec:mod_chem}
\textcolor{green}{\textcolor[rgb]{0,0,0}{On the basis of the dust temperature and the hydrogen density}} maps that were derived using the ray-tracing technique and that are presented in Section~\ref{sec:dustresults}, we model the chemical evolution of the molecular gas in the cloud cores. The density and temperature profiles are kept constant during the chemical evolution of the cores. The gas is initially atomic, comprising 13 elements, and it evolves with time because of the processes described below. This approach of a chemical evolution under constant physical conditions is often called "pseudo time-dependent". We call the time span during which the chemical models evolve "chemical age". How this parameter is related to the "real" age of the cores is discussed in Sect.~\ref{subsec:age}.

We act on the assumption that the gas temperature is equal to the dust 
temperature everywhere in the clouds. This simplifying assumption should 
hold to first order at hydrogen densities above $10^4$\,cm$^{-3}$ where dust and 
gas are collisionally coupled~\citep{galli2002} and where most of the
molecular emission originates. However, the coupling might not be perfect 
even in the densest regions of our study ($>10^6$\,cm$^{-3}$): for example, for B\,68, where we have derived a central hydrogen density of 
$4\times 10^5$\,cm$^{-3}$ and central dust temperature of 8 K, \citet{hotzel2002}
and \citet{lai2003} find a central gas temperature of 10-11~K from 
ammonia observations. On the other hand, ammonia
might be depleted in the core center such that it instead traces a somewhat 
warmer inner layer of the envelope. In our model of B\,68, the dust temperature in
a shell of radius 5000 to 7000 AU is indeed 10-11~K. This could also explain 
at least part of this small discrepancy.
The decoupling of gas and dust temperatures at
lower densities affects our modeling results less than the chemical
evolution, since the rotational transitions at low densities
are subthermally excited. For instance, at a hydrogen density of $10^3$\,cm$^{-3}$ and a
gas temperature of 100\,K, the $J=2-1$ transition of CO would only be
excited to about 9\,K, while the same transition would be excited to 5\,K at a gas temperature of 20\,K\footnote{This has been calculated using the online version of RADEX~\citep{tak2007}}. Therefore, underestimating the gas temperature affects the modeled level populations and the molecular emission at low densities only weakly. 

In the vicinity of the core centers,
the majority of globules show almost circularly symmetric shapes. As we study
these central cloud regions, the modeling is performed assuming 1D radial
profiles. We obtain 1D profiles of the observed molecular emission and the
dust temperature and density maps by azimuthal averaging. We cut out the segment
of the circle in the direction of asymmetries for the averaging task. The regions that have been taken into account for deriving the 1D profiles are indicated in
Figs.~\ref{fig_obs_cb4}~-~\ref{fig_obs_cb244}. The resulting profiles of the dust temperature, hydrogen densities, and column densities are
shown in Fig.~\ref{fig_dust_1D}.

We use the time-dependent gas-grain chemical model `ALCHEMIC'
\citep[see][]{semenov2010}. A brief summary is given below. The chemical
network is based on the 2007 version of the OSU network\footnote{see:
  http://www.physics.ohio-state.edu/$\sim$eric/research.html}. The reaction rates as of November 2010 are implemented (e.g., from the KIDA
database\footnote{http://kida.obs.u-bordeaux1.fr}). We consider cosmic ray
particles (CRP) and CRP-induced secondary UV photons as the only external
ionizing sources. We adopt a CRP ionization rate of $\zeta_{\rm
  CR}=1.3\times 10^{-17}$~s$^{-1}$~\citep{herbst1973}, although recent studies have revealed higher CRP ionization rates in translucent clouds~\citep[e.g.,][]{indriolo2012}. The dense molecular cloud cores studied here are embedded in extended envelopes that, despite their low density, effectively scatter the low energy particles of the cosmic rays and thereby reduce the luminosity of the cosmic rays in the shielded regions. Therefore the lower value of \citet{herbst1973} is expected to hold for dense and shielded clouds~\citep[e.g.,][]{garrod2013,vasyunin2013}. The UV dissociation and ionization
photo rates are adopted from
\citet{dishoeck2006}\footnote{http://www.strw.leidenuniv.nl/$\sim$ewine/photo/}. The self-shielding of H$_2$ from photodissociation is calculated by
Eq. (37) from \citet{draine1996}. The shielding of CO by dust grains,
H$_2$, and its self-shielding are calculated using the precomputed table of
\citet[Table 11]{lee1996}. As the chemical model does not include carbon
isotopologues, the self-shielding factors for $^{12}$C$^{18}$O are calculated
by lowering the $^{12}$C$^{16}$O abundances by an appropriate isotopic ratio
(see below).

In addition to gas-phase chemistry, the chemical model includes the following
gas-grain interaction processes.~(We refer the reader to Sect.~2.3 in \citet{semenov2010} for the corresponding equations.) Molecules accrete on and stick to dust grains by physisorption
with 100\% probability, except for H$_2$, which is not allowed to stick.
Chemisorption of surface molecules is not allowed.  The surface molecules are
released back to the gas phase by thermal, CRP-, and CRP-induced
UV-photodesorption. All these processes are modeled with a first-order
kinetics approach~(Eq. (2) in \citet{semenov2010}).  A UV photodesorption yield for surface species of
$10^{-3}$ is assumed \citep[e.g.,][]{oeberg2009a,oeberg2009b}. The grain
charging is modeled by dissociative recombination and radiative
neutralization of ions on grains, and by electron sticking to grains.  The
synthesis of complex molecules is included, using a set of surface reactions
and photoprocessing of ices from \citet{garrod2006}. We assume that each
$0.1$\,$\mu$m spherical olivine grain provides $\approx 1.88\times 10^6$ surface sites
for accreting gaseous species \citep{biham2001}. The surface recombination
proceeds only via the Langmuir-Hinshelwood mechanism.  Upon a surface
recombination, the reaction products are assumed to remain on the grains.
Following \citet{katz1999}, we use the standard rate equation approach to the
surface chemistry and do not consider tunneling of H and H$_2$ through the
grain surface.  

Overall, our chemical network consists of 656 species made of
13 elements and 7907 reactions.  All chemical parameters are kept fixed
during the iterative fitting of the observational data. With this model we calculate time-dependent CO and
N$_2$H$^+$ abundances and use them for the line radiative transfer modeling and
fitting of the observed spectral maps.

In the cloud 1D physical model, the density profiles are cut off at an outer
radius where either no density could be derived because the continuum emission is too weak or where the morphology of the
cloud becomes asymmetric. These radii are also the outer boundaries for the
chemical modeling. At these radii the density has, however, not yet dropped to
zero as the cores are further embedded in low-density extended envelopes. We therefore have to take into account that the external UV field is already
attenuated at the outer edge of the cores. We derive the
extinction level by converting the column density at the outer boundaries to an
extinction level $A_V$. We use a conversion factor $A_v/N_\mathrm H$ of $1\times
10^{-21}$~cm$^2$. This is the average value estimated by comparing the
N$_\mathrm H$ maps derived from the dust emission to complementary
near-infrared extinction maps \citep[see][]{launhardt2013,kainulainen2006}, which will be presented and discussed in more detail in a forthcoming paper.

We want to compare models and observations of $^{13}$CO, C$^{18}$O, and
$\mathrm{N_2H^+}$. Since the rare isotopologues are not explicitly taken into
account in the chemical modeling, we proceed in two steps. In the first step,
we calculate the chemical model for $^{12}$CO to get the N$_2$H$^+$
abundances accurately. The modeling itself also consists of two steps. First, the
abundance of the molecules and the corresponding shielding factors are
calculated using an analytic approximation. The shielding factors are
subsequently used for the second step, the full time-dependent modeling. To calculate the profiles of $^{13}$CO and C$^{18}$O, we artificially
decrease the $^{12}$CO abundance by a factor in the analytic approximation and
calculate the shielding-factor for these species. They are used for the
time-dependent chemical modeling. The initial abundances are, however, taken to
be the same as for the modeling of $^{12}$CO. To obtain the abundance
of the rare isotopologues, the resulting abundances of the time-dependent
modeling are divided by the corresponding fraction of these species with
respect to their main isotopologue. This approach allows us to take\ the reduced self-shielding of $^{13}$CO and C$^{18}$O into account with
respect to $^{12}$CO without extending the chemical network to the rarer isotopologues. In the literature there is a spread of values mentioned
for the ratios of the isotopologues. We use a fraction of 1/70 for $^{13}$CO
with respect to $^{12}$CO and 1/490 for C$^{18}$O~\citep{wilson1994,lodders2009}.

Since the temperature and abundances of the molecules vary along the
LoS, the molecular emission could be partly
subthermally excited or be optically thick. For these reasons, we calculate the
rotational level populations and emission of the molecules from the synthetic
abundance profiles using line-radiative transfer modeling. It is not our goal to model the kinematics of the cores. For the pure purpose of modeling the lineshape and the total intensity of the transitions, we can neglect variations of macroscopic and turbulent motions in the cores, since they are so weak that this simplification does not affect the results for the molecular abundances. We use the non-LTE line radiative transfer code `LIME'
\citep{brinch2012} and the molecular data files of the ``Leiden Atomic and Molecular Database``~\citep{schoier2005}. The LIME code allows to take into account line-broadening due to micro turbulent motions. This is done by setting a scalar Doppler broadening parameter. Since the linewidths in the globules do not vary significantly over the globules, it is possible to set this parameter such that the linewidths of models and observations agree everywhere. The Doppler broadening parameter is taken as one free parameter in the modeling process. In this way we ensure that the optical depth of the emission is properly taken into account. The macroscopic velocity field in the modeling is set to zero. The velocity channel width is set to the value of the observed spectral cubes and the pixel spacing is set to $10\arcsec$ to which all observational data are also regridded. 

For the comparison of synthetic and observed spectra, the latter have been prepared as follows. We have removed line shifts from macroscopic motions in the observed spectra since we do not model the kinetics of the cores. For this task, the spectra have been fitted and shifted by the derived $\mathrm{v_{LSR}}$ such that the centers of all spectra are located at $\mathrm{v_{LSR}}=0$. They have also been rebinned such that $\mathrm{v_{LSR}}=0$ lies in the center of a bin as is the case for the synthetic spectra. Finally, the spectra are azimuthally averaged. The resulting radially varying spectra build up the reference to which the modeled spectra can be compared directly.

During the course of the modeling, we vary four parameters. The chemical model has only one "true" free parameter, the chemical age, which is the time during which the gas evolves at constant physical conditions. In addition, we take into account uncertainties in the derived hydrogen densities of up to a factor 3 resulting from the uncertainties of the assumed dust properties~(see Sect.~\ref{sec_rt_ana}). The relative initial abundances of the atomic species are kept constant when the hydrogen density is varied. We also need to consider that the hydrogen density has not dropped to zero at the outer boundaries of the model cores. Halos of dust particles around the globules with a mean extinction $A_V$ of typically 1-3\,mag are detected in the SPIRE, as well as NIR extinction maps that are quite extended. Since these envelopes shield the globules from the UV-part of the ISRF, they need to be considered in the chemical modeling. The halos are not strictly spherically symmetric, either in extinction / column density or in extent, as pointed out in \citet{launhardt2013}. To account for this and other uncertainties discussed in Sect.~\ref{sec:uv}, we consider the measured mean envelope extinction as a starting guess only and vary them during the modeling procedure. 

We find the best model in the following way. The relative differences of observed and synthetic spectra are summed up for each velocity bin and at all radii. The model with the lowest value of this sum is taken as the best model. It reproduces not only the integrated emission well, but also the line shape such that the optical depth of the emission is taken into account properly.

\section{Results \& discussion}

\subsection{Dust temperature and density maps}
\label{sec:dustresults}

\begin{figure*}
   \centering
   \includegraphics[width=1.0\textwidth]{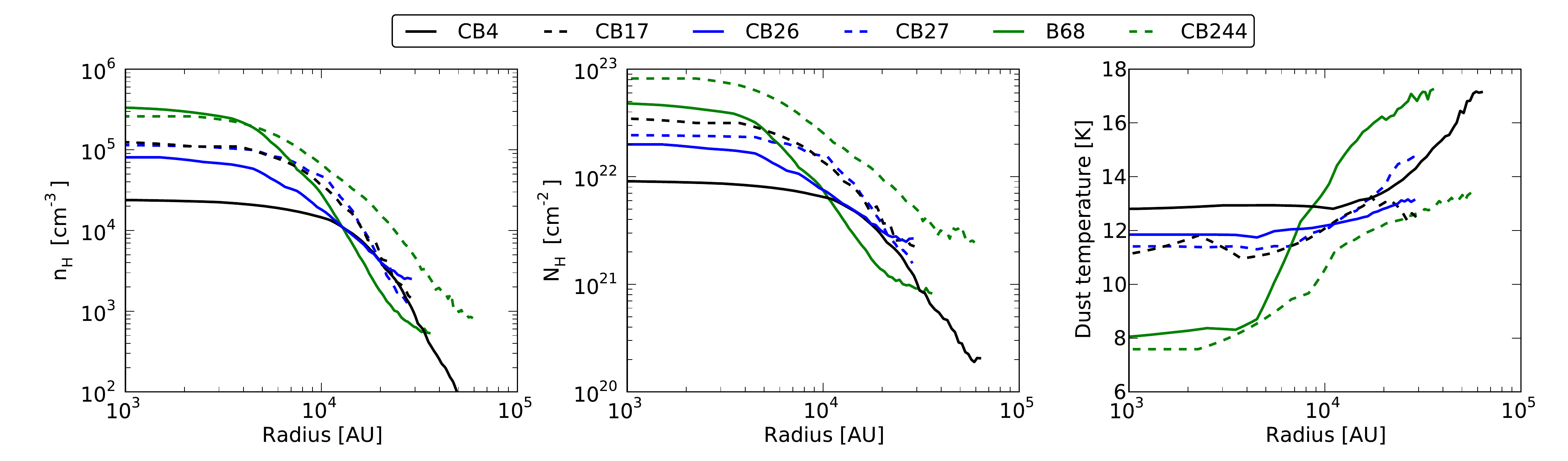}
   \caption{\footnotesize{Azimuthally averaged profiles of the hydrogen density~(left), hydrogen column density~(center), and dust temperature~(right)
     derived with the ray-tracing technique.}}
   \label{fig_dust_1D}
\end{figure*}
The final parameters of the hydrogen number
density and dust temperature profiles in the cores according to the formulae given in
Sect.~\ref{sec_rt_ana} are listed in Table~\ref{tab_rt}. In Figs.~\ref{fig_dustmaps_CB4}~-~\ref{fig_dustmaps_CB244} we show the
hydrogen density and dust temperature maps of all globules in the mid-plane across the sky as they were obtained from the continuum data with the ray-tracing technique described in
Sect.~\ref{sec_rt_ana}. The central density of the starless cores ranges from
$4\times 10^4$\,cm$^{-3}$ to~$4\times 10^5$\,cm$^{-3}$. The cores are clearly non-isothermal, as already predicted by radiative transfer models~\citep[e.g.,][]{evans2001}. In all cores the dust temperature
drops from the outer rims with 13-19~K to the core centers to 8-13~K. Compared
to the temperatures derived from the modified black-body fitting technique in
\citet{launhardt2013}, the central temperatures are lower by
1-3~K. The central column densities reach from $7 \times
10^{21}$\,to~$6\times 10^{22}$\,cm$^{-2}$ in the maps derived with the ray
tracing technique compared to the column densities of $7.5 \times
10^{21}$\,to~$4.6\times 10^{22}$\,cm$^{-2}$ in \citet{launhardt2013} which is on average 60\% more. The lowest difference is found in the case of CB\,17 with an increase of only 6\% and the highest in the case of B\,68 with an increase of 230\%. 

From the maps we have derived one-dimensional profiles by azimuthally
averaging around the centers of the starless cores. The regions that were
considered for deriving the radial profiles are indicated
with thick white lines in the \textit{SPIRE}~250~$\mu$m maps in
Figs.~\ref{fig_obs_cb4}-\ref{fig_obs_cb244}. Because of its strong asymmetry and the nearby protostellar core, the starless core in CB\,130 is not well-suited for a 1D study. We therefore do not derive a radial profile of its density and temperature structure and will also not take it into account for the chemical modeling. The profiles of the hydrogen volume and column density, and dust
temperature of the other globules are plotted in Fig.\ref{fig_dust_1D}.

\begin{table*}
\caption{\label{tab_rt}\footnotesize{Parameters of the hydrogen density and dust temperature profiles corresponding to Equations~(\ref{eq_plummer}) - (\ref{eq_tau})}}
\centering
\begin{tabular}{lccccccccc}
  \hline\hline
  Source & $N_0$\tablefootmark{a} & $n_0$\tablefootmark{a} & $n_{out}$ & $\eta$\tablefootmark{b} & $r_0$ & $r_1$ & $r_{out}$ & $T_\mathrm{in}$\tablefootmark{c} &
  $T_\mathrm{out}$\tablefootmark{c}\\
  &[cm$^{-2}$] & [cm$^{-3}$]&[$10^2$cm$^{-3}$]&&[$10^4$ AU]&[$10^4$ AU]&[$10^4$ AU]&[K]&[K]\\
  \hline
  CB 4 & $9.2\times 10^{21}$ & $2.5\times 10^4$ & 1 & 5.0 & 2.0 & 5.7 & 8 & 13.1 & 19\\
  CB 17\tablefootmark{d} & $3.0\times 10^{22}$ & $1.3\times 10^5$ & 6 &5.0& 1.1 & 3.0 &4&10.5&13\\
  CB 26 & $2.0\times 10^{22}$ & $8.7\times 10^4$ & 5 & 3.0 & 0.8 & 4.4 &3&12.0&14\\
  CB 27 & $2.7\times 10^{22}$ & $1.3\times 10^5$ & 7 & 6.0 & 1.3& 6.6 & 3 &11.3&14\\
  B 68\tablefootmark{d} & $5.5\times 10^{22}$ & $4.0\times 10^5$ & 4  &5.0&0.7& 2.7 & 5&8.1&17\\
  CB 244 & $9.1\times 10^{22}$ & $2.5\times 10^
5$ & 7 & 4.0 & 1.1 & 4.7 & 7 & 8.4 & 14\\
  \hline
\end{tabular}
\tablefoot{
\tablefoottext{a}{The estimated relative uncertainty is $\pm 25$\%, 
 plus a scaling uncertainty by up to a factor of 3 due to the not well-constrained dust 
 opacity model (see uncertainty discussion in Sect. 3.1.).}
\tablefoottext{b}{The estimated uncertainty is $\pm 1$.}
\tablefoottext{c}{The estimated uncertainty is $\pm 1$\,K. The temperature estimate is 
 less affected by the dust model (see Launhardt et al. 2013).}
\tablefoottext{d}{Slight differences to the values derived by Schmalzl et al.~(submitted to A\&A) 
 and Nielbock et al. (2012) can be accounted to the uncertainties discussed in Sect. 3.1.).}
}
\end{table*}

\subsection{Sizes, masses, and stability of the cores.}
\label{sec:stability}

\begin{figure}
   \centering
   \includegraphics[width=0.35\textwidth]{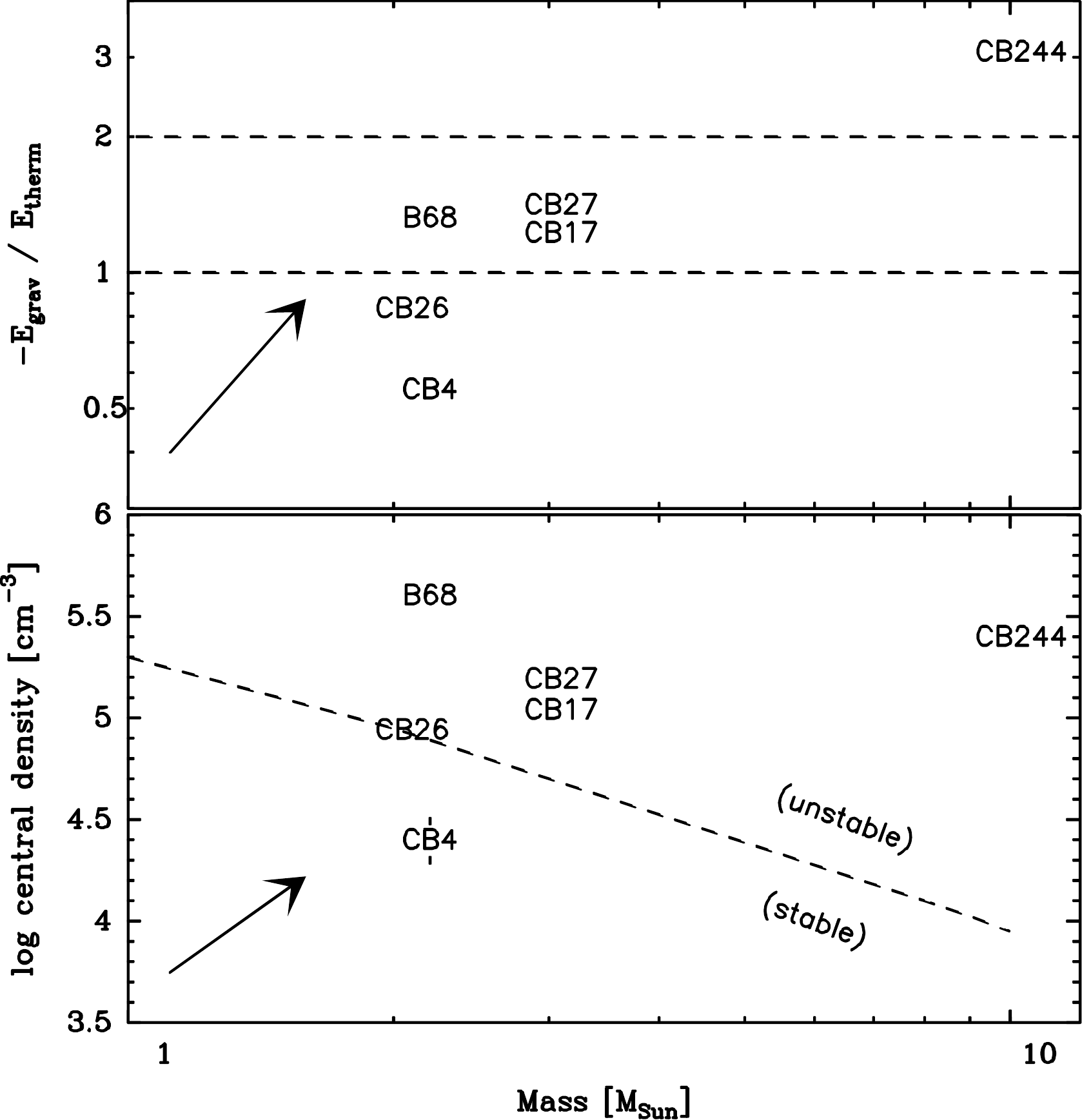}
   \caption{\footnotesize{Stability of the cores. Same plot as Fig.~6 in
     \citet{launhardt2013}, but here for the results of the ray-tracing
     models. \textit{Top}: ratio of gravitational potential to thermal kinetic
     energy vs. total gas mass ($M_\mathrm{core}$). The lower dashed
     horizontal line marks the bounding limit of
     $-E_\mathrm{grav}/E_\mathrm{therm}=1$, while the upper line marks the
     state of virialization at
     $-E_\mathrm{grav}/E_\mathrm{therm}=2$. \textit{Bottom}:~Central density
     vs. total gas mass for the same cores. The dashed line marks the maximum
     stable density of a pressure-supported, self-gravitating modified
     (nonisothermal) BES as calculated by \citet[][with their Fig.~14, model with
     photoelectric heating at the core boundary taken into
     account]{keto2008}. The uncertainty resulting from the ray-tracing modeling is indicated by the error bars at CB\,4. The arrows indicate the maximum systematic shift of the cores in the diagram if grain properties of ISM dust would be used in the modeling.}}
   \label{fig_stability}
\end{figure}

We derived masses of the starless cores by integrating over the circularized 1D hydrogen
density profiles~$n_\mathrm H(r)$. We conservatively
estimate the mass by choosing $r_1$\,(Eq.~\ref{eqr1}) as outer radii of the cores
where the profiles turn from a power-law like into a
flat shape. These radii range from $27\times 10^3$\,to~$57\times 10^3$\,AU and are
listed in Table~\ref{tab_rt}. We take into account that the total gas mass
is higher by a factor of $\mu_\mathrm H=1.36$ than the hydrogen atom mass so that the core
masses are given by:
\begin{equation}
M_\mathrm{core}=M(r_1)=\mu_\mathrm H m_\mathrm H \times 4\pi\int_{0}^{r_1}n_\mathrm{H}(r)r^2\mathrm dr
\end{equation}
The resulting core masses range from 2~M$_{\sun}$ to 10~M$_{\sun}$\,(see
Fig.~\ref{fig_stability}) and are well in agreement with those previously
estimated from modified black-body fits~\citep[see][]{stutz2010,launhardt2013}.

The stability of the cores was checked by comparing the gravitational
potential with the thermal energy of the particles in the cores. We neglected turbulent pressure, rotational energy and magnetic fields. This is justified since the thermal energy is the strongest contributor to the stabilization. Assuming typical magnetic field strengths in Bok globules of a few 100\,$\mu$G~\citep{wolf2003}, these three contributions are dominated by the turbulent pressure which we have estimated for all cores and typically adds only 10\% to the thermal energy.

We estimated the gravitational potential by

\begin{equation}
E_\mathrm{grav}=-4\pi G m_\mathrm H \mu_\mathrm H \int_{0}^{r_1}M(r)n_\mathrm H(r)r\mathrm dr
\end{equation}
where G is the gravitational constant. The
thermal energy of the molecules is calculated from
\begin{equation}
E_\mathrm{therm}=\frac{3k_\mathrm B \mu_\mathrm H}{2\mu}\,4\pi\int_{0}^{r_1}T(r)n_\mathrm H(r)r^2\mathrm d r
\end{equation}
where $k_\mathrm B$\,is the Boltzmann constant and $\mu_\mathrm H n_\mathrm H/\mu$ with $\mu/\mu_\mathrm H=0.6$ the mean particle density for molecular gas. For the gas temperature, we
adopt the dust temperature. $T_\mathrm{Gas}$ might actually be higher than $T_\mathrm{Dust}$ in the outer regions of the envelopes, so that we probably underestimate $E_\mathrm{therm}$.

In the top panel of Fig.~\ref{fig_stability}, we plot the ratio of the total
gravitational to thermal energy of all starless cores against their masses. The
dashed horizontal line at a ratio of 2 indicates virial equilibrium while the
dashed line at a ratio of 1 marks the bounding limit. The values for the cores indicate that
only the starless core of CB\,244 is clearly gravitationally unstable.  In
Fig.~\ref{fig_stability}~(bottom) we plot the central density against the
total mass of the cores. The dashed line indicates the maximum central
density of a pressure-supported, self-gravitating modified (non-isothermal)
Bonnor-Ebert sphere \citep[BES;][]{bonnor1956,ebert1955}, calculated by
\citet{keto2008} for a core with photoelectric heating at the boundary. The
comparison of the core properties to the theoretical curve strengthens the
result that the starless core in CB\,244 is prestellar. It also suggests that CB\,17, CB\,27, and B\,68 are thermally super critical.

\subsection{Spectra and molecular emission maps}

\begin{table*}
\caption{\label{tobsResults}\footnotesize{Spectral line parameters at the centers of the
  starless cores}}
\centering
\begin{tabular}{lccccccccc}
  \hline\hline
  Source & Line & $\Delta \mathrm{v}_{\rm chann}$ & $\mathrm{v}_{\rm LSR}$\tablefootmark{(a,b)} & $T_{\rm mb}^{\rm
    peak}$\tablefootmark{(a,b)}& $\Delta T_{\rm mb}^{\rm RMS}$ & $\Delta \mathrm{v}_{\rm FWHM}$\tablefootmark{(a,b)} & $\Delta \mathrm{v}_\mathrm{therm}$\tablefootmark{(c)} & $\Delta \mathrm{v}_\mathrm{non-therm}$& $I_{\rm line}$\tablefootmark{(d)}\\
  && [m/s] & [km/s] & [K] & [K] & [km/s] & [km/s]& [km/s]& [K.km/s] \\
  \hline
  CB 4 & $^{12}$CO(2-1) & 0.325 & $-11.423\pm 0.002$ & $13.3\pm 0.3$ & 0.11 &
  $0.510\pm 0.007$ & 0.15& $0.49\pm0.01$  & $7.8\pm0.6$\\
  & $^{13}$CO(2-1)  & 0.34 & $-11.35\pm 0.01$ & $7.3\pm 0.6$ & 0.22 & $0.45\pm 0.02$ &
  0.15 & $0.42\pm0.03$ & $4.4\pm0.3$\\
  & C$^{18}$O(2-1)  & 0.053 & $-11.442\pm 0.004$ & $3.5\pm 0.3$ & 0.15 & $0.19\pm 0.01$ & 0.15 & $0.12\pm0.04$ & $0.73\pm0.04$ \\
  & N$_2$H$^+$(1-0) & 0.063 & $-11.355\pm 0.005$ & $0.47\pm 0.09$  & 0.09 & $0.17\pm 0.02$& 0.15 & $0.08\pm0.08$ & $0.38\pm0.08$\\
  \hline
  CB 17 & $^{12}$CO(2-1) & 0.325 & $-4.74\pm0.01$ &  $9.7\pm0.5$ & 0.15 & $0.97\pm0.03$ & 0.14 & $0.96\pm0.03$ & $12.8\pm0.8$\\
  & $^{13}$CO(2-1) & 0.34 & $-4.70\pm0.01$ & $6.2\pm0.4$ & 0.19 & $0.71\pm0.03$ & 0.14 & $0.70\pm0.03$ & $2.5\pm0.4$\\
  & C$^{18}$O(2-1) & 0.053 & $-4.764\pm0.004$ & $2.4\pm0.1$ &0.07& $0.45\pm0.01$ & 0.13 & $0.43\pm0.02$ & 1.2\\
  & N$_2$H$^+$(1-0) & 0.063 & $-4.630\pm0.001$ & $1.5\pm0.2$ & 0.19 & $0.325\pm0.002$ & 0.14 & $0.29\pm0.01$ & 3.0\\
  \hline
  CB 26 & $^{12}$CO(2-1) & 0.325 & $+5.590\pm 0.007$ & $3.6\pm0.1$& 0.08 & $1.17\pm0.02$ & 0.14 & $1.16\pm0.02$ 
  & $6.3\pm0.3$\\
  & $^{13}$CO(2-1) & 0.34 & $+5.64\pm0.01$ & $2.7\pm0.2$ & 0.14 & $1.01\pm0.03$ & 0.14 & $1.00\pm0.03$ & $2.3\pm0.5$\\
  & C$^{18}$O(2-1) & 0.053 & $+5.60\pm0.02$ & $1.1\pm0.1$& 0.15 & $0.70\pm0.04$\tablefootmark{(e)}& 0.14 & $0.69\pm0.04$ & $0.73\pm0.09$\\
  & N$_2$H$^+$(1-0) & 0.063 & \tablefootmark{(f)} & \tablefootmark{(f)} &  0.08 &\tablefootmark{(f)}&\tablefootmark{(f)}&\tablefootmark{(f)} & $0.4\pm0.1$ \\
  \hline
  CB 27 & $^{12}$CO(2-1) & 0.325 & $+7.127\pm0.005$ & $4.1\pm0.1$ & 0.08 & $0.89\pm0.01$ & 0.14 & $0.88\pm0.01$ & $4.4\pm0.2$\\
  & $^{13}$CO(2-1) & 0.34 & $+7.14\pm0.02$ & $3.4\pm0.05$ & 0.10 & $0.61\pm0.05$ & 0.14 & $0.59\pm0.06$ & $2.2\pm0.3$\\
  & C$^{18}$O(2-1) & 0.053 & $+7.058\pm0.005$ & $3.0\pm0.2$ & 0.16 & $0.24\pm0.01$ & 0.13 & $0.20\pm0.02$ & $0.78\pm0.06$\\
  & N$_2$H$^+$(1-0) & 0.063 & $+7.104\pm0.001$ & $2.1\pm0.08$ & 0.08 & $0.187\pm0.004$ & 0.14 & $0.12\pm0.03$ & $2.15\pm0.07$\\
  \hline
  B 68 &$^{12}$CO(2-1) & 0.325 & $+3.30\pm0.01$ & $5.9\pm0.4$ & 0.16 & $0.63\pm0.02$ & 0.14 & $0.61\pm0.03$ & $4.4\pm0.3$\\
  & $^{13}$CO(2-1) & 0.34 & $+3.35\pm0.01$ & $3.1\pm0.5$ & 0.12 & $0.51\pm0.05$ & 0.14 & $0.49\pm0.06$ & $1.8\pm0.3$\\
  & C$^{18}$O(2-1) & 0.030 & $+3.292\pm0.001$ & $2.10\pm0.02$ & 0.02 &
  $0.200\pm0.001$ & 0.14 & 0.$14\pm0.02$ & $0.45\pm0.04$\\
  & N$_2$H$^+$(1-0) & 0.063 & $+3.347\pm0.001$& $0.88\pm0.01$ & 0.01 & $0.249\pm0.003$ & 0.14 & $0.21\pm0.02$ & $1.5\pm0.04$\\
  \hline
  CB 130 & $^{12}$CO(2-1) & 0.051 & \tablefootmark{(g)}& 2.3\tablefootmark{(h)} & 0.08 & 4.2\tablefootmark{(i)} & 0.14 & ..& $11.6\pm0.1$\\
  & $^{13}$CO(2-1) & 0.053 & $+7.60\pm0.01$ & $1.2\pm0.1$ & 0.07 & $1.26\pm0.04$\tablefootmark{(e)} & 0.14 & $1.25\pm0.04$ & $1.8\pm0.1$\\
  & C$^{18}$O(2-1) & 0.053 & $+7.524\pm0.006$ & $0.96\pm0.05$ & 0.05 & $0.48\pm0.01$ & 0.14 & $0.46\pm0.02$ & $0.52\pm0.03$\\
  & N$_2$H$^+$(1-0) & 0.063 & $+7.514\pm0.002$ & $1.34\pm0.04$ & 0.04 & $0.397\pm0.005$ & 0.14 & $0.37\pm0.01$ & $3.2\pm0.05$\\
  \hline
  CB 244 & $^{12}$CO(2-1) & 0.325 & $+3.407\pm0.008$ & $4.2\pm0.1$ & 0.05 & $1.62\pm0.02$ & 0.13 & $1.61\pm0.02$ & $7.5\pm0.5$\\
  & $^{13}$CO(2-1)& 0.34 & $+3.86\pm0.01$ & $2.5\pm0.1$ & 0.07 & $0.98\pm0.03$ & 0.13 & $0.97\pm0.03$ & $2.7\pm0.2$\\
  & C$^{18}$O(2-1) & 0.053 & $+3.805\pm0.007$ & $3.0\pm0.2$ & 0.19 & $0.43\pm0.02$ & 0.13 & $0.41\pm0.03$ & $1.2\pm0.1$\\
  & N$_2$H$^+$(1-0) & 0.063 & $+3.875\pm0.002$& $1.90\pm0.08$ & 0.08 & $0.34\pm0.005$ & 0.13 & $0.31\pm0.01$ & $4.6\pm0.2$\\
  \hline\hline
\end{tabular}
\tablefoot{
  \tablefoottext{a}{for the CO isotopologues determined from Gaussian fits, for N$_2$H$^+$ determined from fits of the hyperfine structure,}
  \tablefoottext{b}{for N$_2$H$^+$ of the main line at 93173.777\,MHz,}
  \tablefoottext{c}{assuming a constant gas temperature along the LoS that is equal to the dust temperature derived from modified black body fitting. We estimate an uncertainty 0.02km/s,}
   \tablefoottext{d}{Determined by integrating of the spectra (in the case of N$_2$H$^+$ all components),}
  \tablefoottext{e}{This line is double peaked. The linewidth given here is given for a Gaussian fitted to the full spectrum,}
  \tablefoottext{f}{not detected,}
  \tablefoottext{g}{not determinable,}
  \tablefoottext{h}{not fit with a Gaussian. Peak value taken directly from the spectrum.,}
  \tablefoottext{i}{not fit with a Gaussian. Width of spectrum where $T_\mathrm{mb}>0.5 T_\mathrm{mb}^\mathrm{peak}$.}
}
\end{table*}

\begin{table*}
\caption{\label{tabCO}\footnotesize{Ratio of icy to total C$^{18}$O abundance abundance in the core centers, radius, hydrogen column density, hydrogen density, and dust temperature at which 50\% of the total C$^{18}$O abundance is frozen onto the grains.}}
\centering
\begin{tabular}{lccccc}
  \hline\hline
&$X$(sC$^{18}$O)/&R(50\%)&$N_\mathrm H$(50\%)&$n_\mathrm H$(50\%)&$T_\mathrm D$(50\%)\\
& $X$(C$^{18}$O) [\%] &[AU]&[cm$^{-2}$]&[cm$^{-3}$]&[K]\\
\hline
CB\,4 & 60 & 6,000 & $7.7\times 10^{21}$ & $5.7\times 10^4$ & 12.9 \\
CB\,17 & 77 & 9,000 & $1.6\times 10^{22}$& $1.5\times 10^5$ & 11.1 \\
CB\,26 & 48 & ... & ... & ... & ... \\ 
CB 27 & 98 & 12,000 & $1.1\times 10^{22}$& $8.2\times 10^4$ & 12.5 \\
B 68 & 98 & 7,000 & $1.8\times 10^{22}$ & $8.9\times 10^4$& 10.5 \\
CB 244 & 95 & 10,000& $2.8\times 10^{22}$& $1.6\times 10^5$ & 9.8\\
\hline
\end{tabular}
\end{table*}

\begin{figure*}
   \centering
   \includegraphics[width=0.85\textwidth]{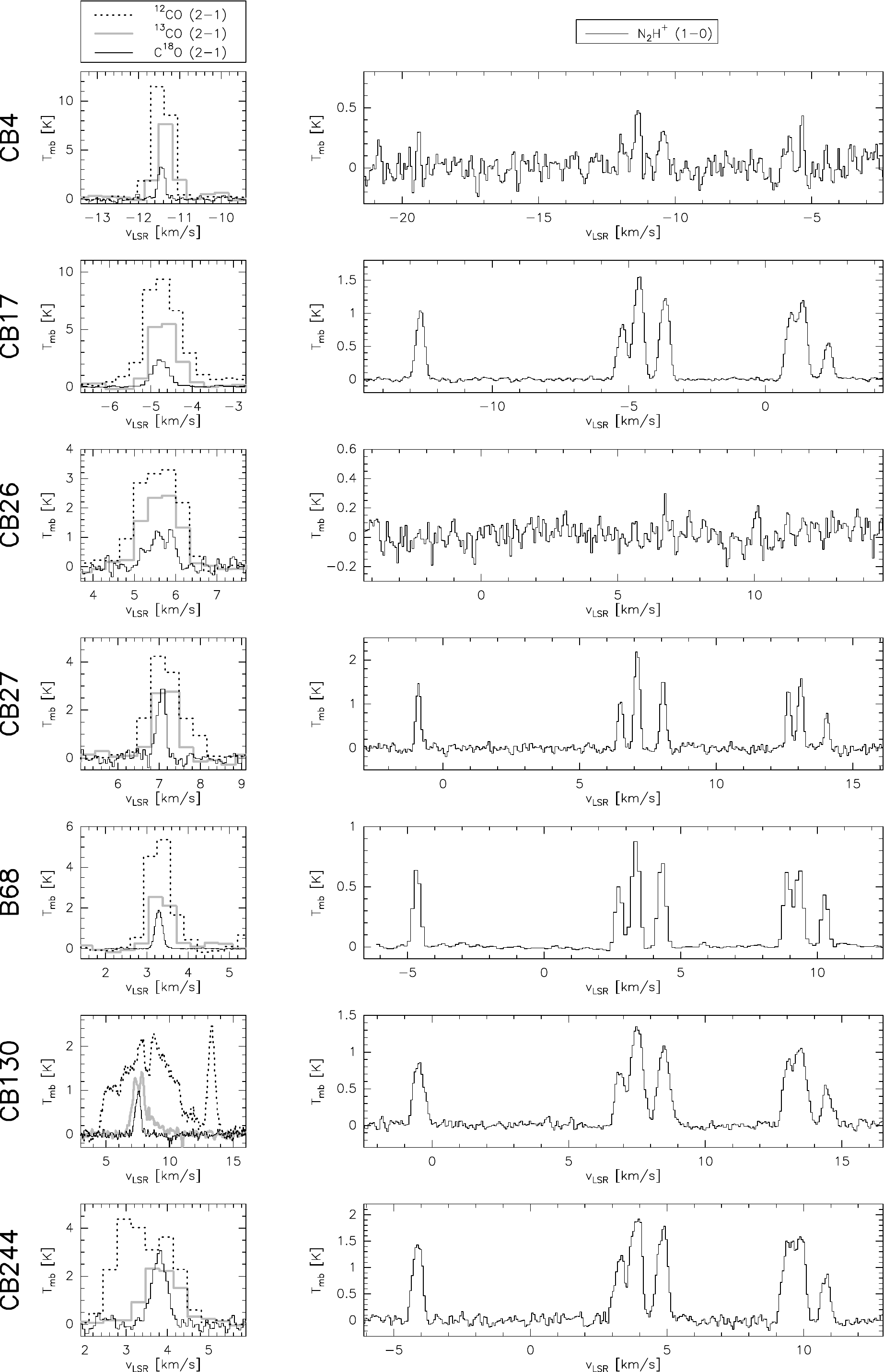}
   \caption{\footnotesize{Spectra of the $^{12}$CO\,(2-1), $^{13}$CO\,(2-1), C$^{18}$O\,(2-1) and N$_2$H$^+$\,(1-0) transitions in a 36.4$\arcsec$ HPB at the centers of the starless cores.}}
   \label{fig_spectra}
\end{figure*}

In Figs.~\ref{fig_obs_cb4}-\ref{fig_obs_cb244}, integrated emission maps of
$^{12}$CO\,(J=2-1), $^{13}$CO\,(J=2-1), C$^{18}$O\,(J=2-1) and N$_2$H$^+$\,(J=1-0) are
presented. For the sake of comparison we also have included maps showing the same regions
as observed with the SPIRE bolometer in the 250 $\mu$m band and from the
Digitized Sky Survey 2 archive\footnote{http://archive.stsci.edu/cgi-bin/dss\_form}. Because of their different abundances, the emission of three isotopologues becomes optically thick at different column densities. As the main CO isotopologue, $^{12}$CO traces the envelopes of the globules particularly well, while $^{13}$CO is best suited to
tracing the outer parts of the cores. As the rarest isotopologue, C$^{18}$O traces the
dense regions of the cores up to hydrogen densities of $\sim 1\times
10^5$\,cm$^{-3}$. At these densities CO freezes out onto the dust
grains. We therefore complemented our set of molecular line observations with
maps of N$_2$H$^+$, which traces the densest parts of the cores where CO is
frozen out. The time scales of build-up and depletion of N$_2$H$^+$ are longer
than for CO so that both species in combination allow constraining the ages of
the cores.

The $^{12}$CO and $^{13}$CO maps are smaller than the \textit{SPIRE 250}\,$\mu$m maps,
but the molecular emission of $^{12}$CO is more extended than the continuum emission in some cases. This
indicates that there are extended cold envelopes to which the
continuum observations are less sensitive. C$^{18}$O is the least abundant
CO isotopologue used in this study. It is therefore the optically thinnest and best suited CO isotopologue for the dense regions of the cores where CO is not frozen out. While its emission is enhanced
at the positions of some Class I objects, it does not show enhancement at the
locations of the starless cores. In contrast, the N$_2$H$^+$ emission shows a
positive gradient towards the core centers. In the two least dense starless
cores, CB\,4 and CB\,26, N$_2$H$^+$ emission was only marginally detected.
  
The spectral line parameters at the centers of the starless cores are listed
in Table~\ref{tobsResults}. The spectra of all molecules toward the centers of the
starless cores are also plotted in Fig.~\ref{fig_spectra}. C$^{18}$O is single-peaked in
all sources except CB\,26, where two distinct peaks at a velocity separation
of 0.33\,km/s are observed. It is unlikely that the dip is due to self-absorption, since the line is quite weak. It is more likely caused by multiple velocity components. Almost no N$_2$H$^+$ emission is detected from the starless core in CB\,26. At the position of the protostar in CB\,26, the
C$^{18}$O line is also double peaked, with the dip at the same velocity as for the
starless core. This is an indication of the protoplanetary disk seen edge-on~\citep{launhardt2001}. The peak main beam temperature
$T_\mathrm{MB}$ of the C$^{18}$O $(J=1-0)$\,transition ranges from 2 K to 6\,K in
the various globules.

The channel width of the $^{13}$CO observations is only 0.34~km/s compared
to 0.05~km/s of the C$^{18}$O spectra. If the $^{13}$CO line of CB\,26 was double
peaked as the C$^{18}$O line, we could not resolve it. The $^{13}$CO line is
single peaked in all cores except for CB\,130 where it shows two peaks with a
velocity separation of 0.53\,km/s. The dip of this line falls into the peak of
the C$^{18}$O emission and is thus probably a result of self-absorption.

The emission of the main isotopologue $^{12}$CO is optically thick
towards the core centers but traces also the outskirts of the globules
where the emission of the less abundant isotopes drops below the detection
limit. Consequently, its linewidth is highest throughout the sample and ranges
from 0.3 km/s to 0.7 km/s in the globules that only contain starless
cores. In those globules that also inhabit protostellar cores, the
linewidths are generally higher and reach values of up to 1.6 km/s. CB\,130 is
an exceptional case with a very broad line emission, consistent with its location within a larger diffuse structure.

All globules show signs of slow large-scale motions. The linewidths are typically larger than the thermal linewidths and
increase from the rare to the abundant CO isotopologues. The average
linewidths of the spectra toward the core centers are 0.30\,km/s, 0.69\,km/s, and
1.0\,km/s for C$^{18}$O, $^{13}$CO, and $^{12}$CO, respectively. The thermal
linewidths are typically around 0.15\,km/s, if one assumes that the gas
temperature is equal to the dust temperature. The radiation of $^{13}$CO and $^{12}$CO
emanate mainly from regions with $n_\mathrm H \lesssim 10^4$\,cm$^{-3}$ where the gas is expected to be warmer than the
dust, since the density is too low to couple both temperatures via
collisions. Therefore, a fraction of the linewidth increase can be attributed to higher
gas temperatures at the outskirts of the globules. Gas temperatures of 30\,K to 50\,K would, however, only result in a thermal line broadening of 0.22\,km/s to 0.29\,km/s, respectively. The remaining broadening must be caused by turbulent or macroscopic motions, but is partly also mimicked by the optical thickness of the lines. 

With the N$_2$H$^+$ observations we trace the inner part ($r \lesssim
10^4$\,AU) of the cores where large amounts of CO are expected to be frozen out
onto the dust grains. The spectra at the position of the core centers are
shown in Fig.~\ref{fig_spectra}~(right panels). The linewidths range
from 0.08~km/s up to 0.37~km/s. This range can only partially result from different temperatures and must have its origin in the different dynamics of the gas.

\subsection{Core properties constrained from chemical modeling}
\label{sec:results_modeling}
\begin{figure*}
   \centering
   \includegraphics[width=0.98\textwidth]{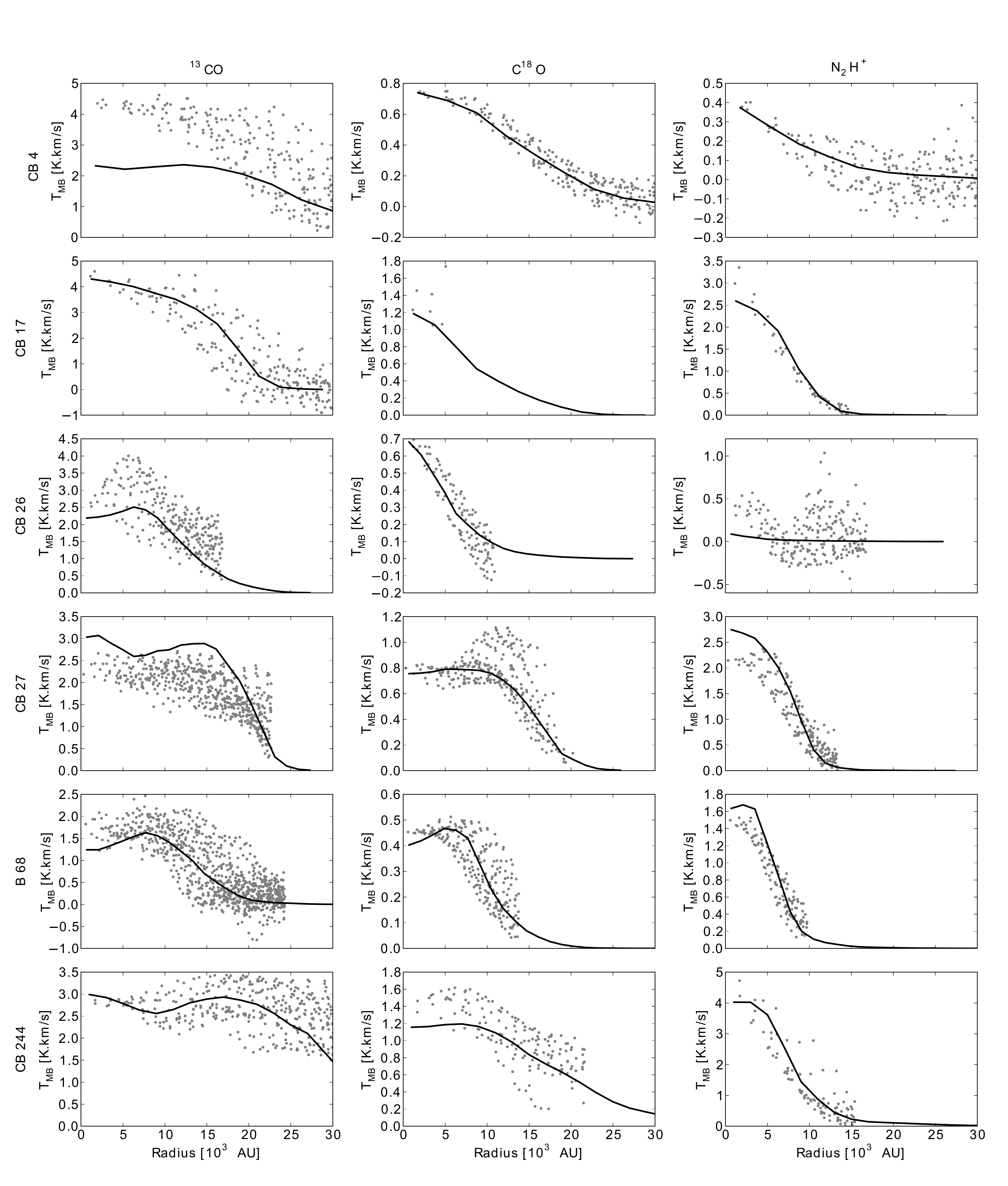}
   \caption{\footnotesize{Profiles of the integrated emission that would result from the molecular abundances of the final models (solid line) compared to the observed profiles of the integrated emission for all modeled starless cores. (Left)  $^{13}$CO~(J=2-1), (center) C$^{18}$O~(J=2-1), (right) N$_2$H$^+$\,(J=1-0).}}
   \label{fig_mod_bestfits}
\end{figure*}

\begin{figure*}
   \centering
   \includegraphics[width=0.98\textwidth]{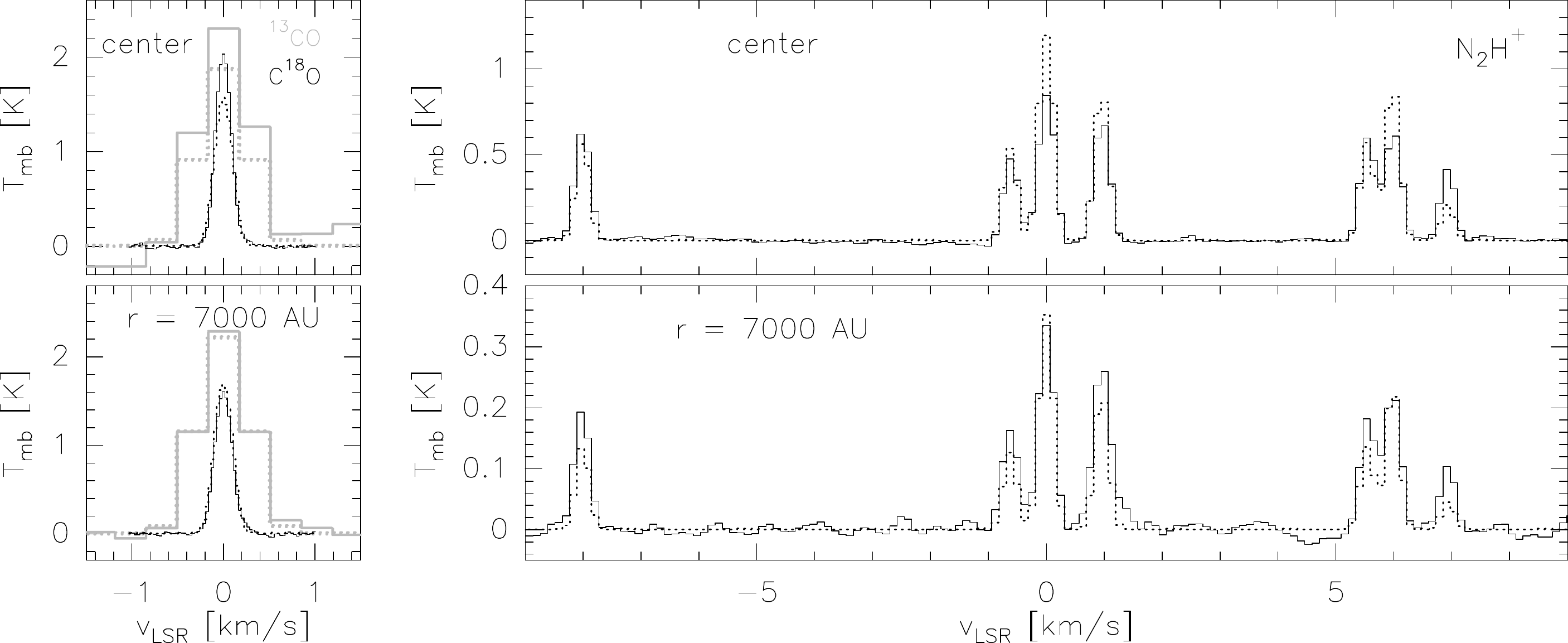}
   \caption{\footnotesize{To quantify the agreement of models and observations we compare the spectra over the full radius of the models. Here, an example of B\,68 is shown. The solid lines represent observations, the dotted lines the synthetic spectra of the best models. On the lefthand side, we show the spectra of the $^{13}$CO~(J=2-1) (gray) and C$^{18}$O~(J=2-1) (black) transitions. On the righthand side, we show the spectra of the N$_2$H$^+$\,(J=1-0) transition. The top line shows the spectra at the position of the core center, the bottom line the spectra at a radial distance of 7000\,AU.}}
   \label{fig_compare_obs_mod}
\end{figure*}

\begin{figure*}
   \centering
   \includegraphics[width=0.98\textwidth]{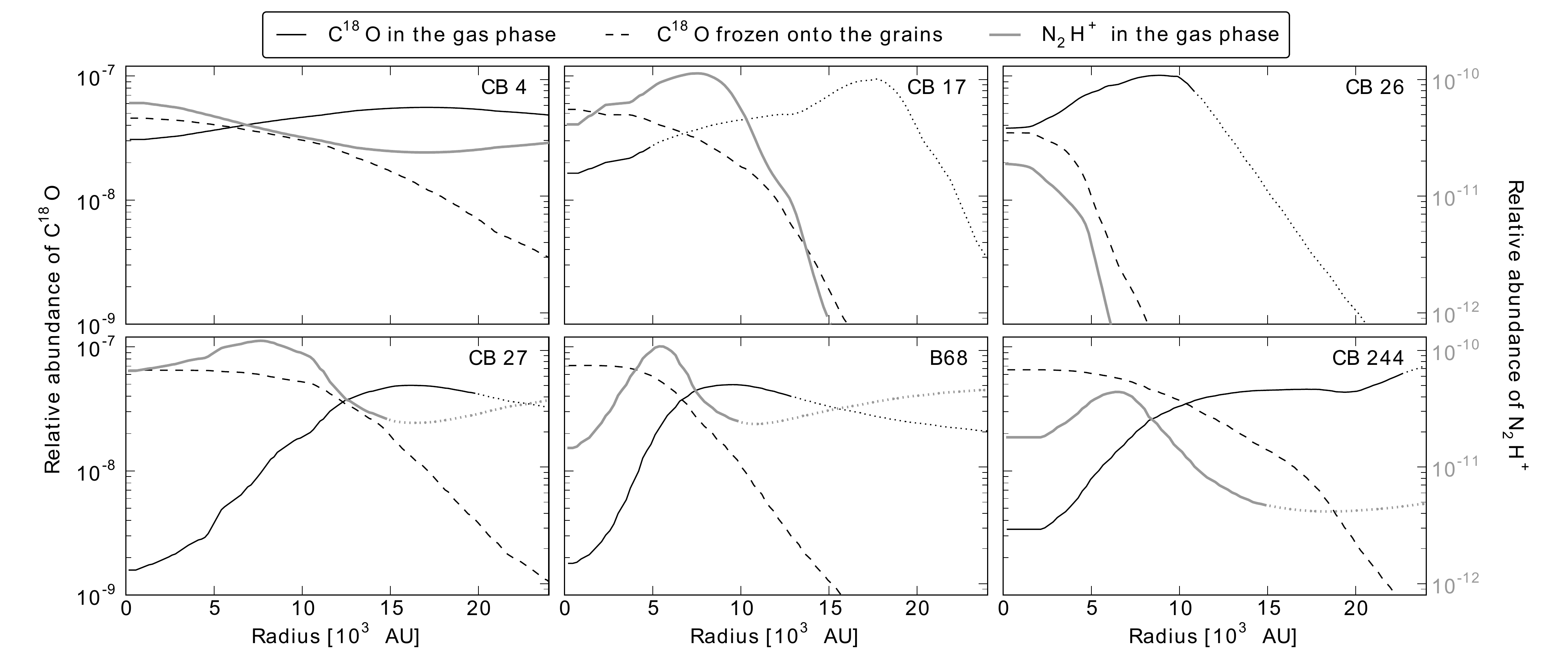}
   \caption{\footnotesize{Relative abundances (absolute molecular density with respect to the hydrogen density) of gaseous C$^{18}$O~(black solid line), frozen-out C$^{18}$O~(dashed black line) and gaseous N$_2$H$^+$ (gray solid line) in the final models. At the outer radii of the emission profiles derived from the observations, the lines turn from a solid into a dotted line shape, indicating that these parts of the profiles are not constrained by observations.}}
   \label{fig_abundances}
\end{figure*}

On the basis of the dust temperature and hydrogen density profiles presented in Sect.~\ref{sec:dustresults}, we have modeled the chemical evolution of the globules. The approach is explained in Sect.~\ref{sec:mod_chem}. The best fits of the modeled integrated emission to the observed profiles are presented in Fig.~\ref{fig_mod_bestfits}. In Fig.~\ref{fig_compare_obs_mod}, we demonstrate the quality of the agreement between modeled and observed line shapes.

\textcolor{green}{\textcolor[rgb]{0,0,0}{\subsubsection{The individual globules}}}

\textbf{CB\,4:} While CB\,4 is the most symmetric and isolated globule of our
study~(Fig.~\ref{fig_obs_cb4}), it is also the least dense one with a central hydrogen density of only $4
\times 10^4$\,cm$^{-3}$. It is the only core in the study for which the modeled
$^{13}$CO emission is underestimated with all parameter sets. Owing to its low density, the coupling of gas and dust temperature is
expected to be the weakest in this globule and the gas temperature might thus be significantly higher than the dust temperature. The emission of C$^{18}$O and N$_2$H$^+$ is
dominated by inner regions such that the gas temperature is closer to the dust
temperature. We find that the chemical age of CB\,4 is $\sim 10^5$\,yr. The significance of the derived chemical ages is discussed in Sect.~\ref{subsec:age}. To match the observations, we have to increase the hydrogen density by a
factor of 3. We require $A_{V,~\mathrm{out}}$ to be set at a value of at least 4~mag,
while a value of 0.6~mag has been derived from the NIR~extinction maps.  

\textbf{CB\,17:} The $^{13}$CO emission of CB\,17 can be reproduced assuming a
chemical age of $7\times 10^4-10^5$\,yr and assuming that the hydrogen density
is increased by a factor of 3. We only have a very small map of
C$^{18}$O emission for this source, such that the observational constraint to the model is limited. We find the best fit for an age of $5\times 10^4-10^5$\,yr. The best model for N$_2$H$^+$ is found at a chemical age of $10^5$\,yr. We find the best fits for an external
visual extinction of 2~mag that is also found from the NIR extinction mapping, but
the deviation is not strong for $A_{V,~\mathrm{out}}$~values ranging from 0 to 6~mag. Our finding of a low chemical age does not confirm the finding of \citet{pavlyuchenkov2006} who found a chemical age of 2\,Myr for the starless core in CB\,17.

\textbf{CB\,26:} We reproduce the $^{13}$CO~emission from the starless core in CB\,26 with a chemical age of $4\times 10^5$\,yr. The emission profile of C$^{18}$O, however, is very steep and shows no signatures of freezeout. It is best described with chemical models at an age of only $\sim 10^4$\,yr and by setting the external visual extinction at the model boundary to 0 ($A_{V,~\mathrm{out}}=1$ in the NIR extinction map). The parameters are consistent with the very weak observed emission of N$_2$H$^+$. 

\textbf{CB\,27:} The $^{13}$CO~emission of CB\,27 cannot be fit well. In particular, the emission coming from large radii is underestimated in all our models. This discrepancy can be overcome if an increased gas temperature in the envelope of this core is assumed. The range of chemical ages, for which the models fit the data similarly well within the uncertainties, is wide and reaches from $10^5$~to~$10^6$\,yr. For the C$^{18}$O~emission, we find the best models at $A_{V,~\mathrm{out}}$~values ranging from 2 to 6~mag ($A_{V,~\mathrm{out}}=1$ has been found in the NIR-extinction map) and an age between $5 \times 10^4$ and $10^5$\,yr. The hydrogen density needs to be multiplied by a factor of 3 with respect to the density profiles obtained from the EPoS project. The N$_2$H$^+$~emission is modeled best with the same age as the C$^{18}$O~emission but only a factor 2 in density.

\textbf{B\,68:} This is the only core for which all three molecular emission profiles can be reproduced from our chemical models with the hydrogen density profile as derived in Sect.~\ref{sec:dustresults}. If we increase the hydrogen density in the models, the resulting molecular emission profiles deviate significantly from the observed profiles. The best chemical age to match the $^{13}$CO observations is $2\times 10^5$\,yr, while we find that the C$^{18}$O and N$_2$H$^+$~profiles are best explained by models with a chemical age between $7\times 10^4$~and~$10^5$\,yr. \citet{bergin2006} found an chemical age of $\sim 10^5$\,yr from chemical modeling and a comparison of the modeled and observed $^{13}$CO and C$^{18}$O emission. This is well in agreement with our findings, and $A_{V,~\mathrm{out}}$ needs to be around 4~-~6~mag in the models while we derived a value of about 1~mag from the NIR~extinction map.

\textbf{CB\,244:} As for CB\,27, all our models underestimate the $^{13}$CO~emission at large radii, which could be explained by a decoupling of dust and gas temperatures at low densities. The chemical age that fits the data best is around $2\times 10^5$\,yr. In contrast, the C$^{18}$O and N$_2$H$^+$~profiles are best fit with a chemical age of only $4\times 10^4$\,yr. Changing the external $A_v$ at the model boundaries does not have a significant impact on the goodness of the fits. The hydrogen density needs to be multiplied by 2 for the CO isotopologues and by a factor of 3 for N$_2$H$^+$.

That we obtain a discrepancy of a factor~$\sim 2$-3 is an indication that the assumed dust model may be incorrect. This result is discussed in Sect.~\ref{sec:disc_grain}.

\subsubsection{CO abundances}

\begin{figure*}
   \centering
   \includegraphics[width=1\textwidth]{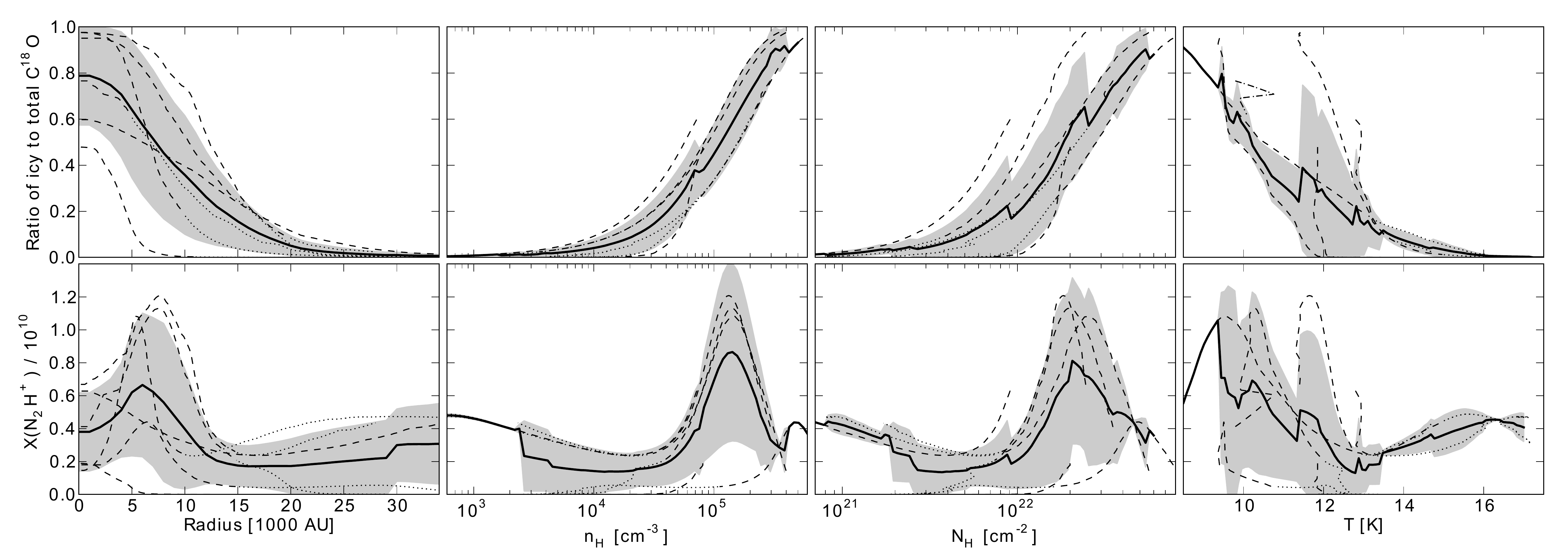}
   \caption{\footnotesize{\textit{Top row}: The ratio of icy-to-total C$^{18}$O abundance of the best-fit models is plotted against the core radius, the hydrogen number density, the column density, and the dust temperature. \textit{Bottom row}: The N$_2$H$^+$ abundance (relative to the hydrogen density) is plotted against the same quantities. The dashed lines indicate the results of the individual globules. Where the models cannot be compared to observations (because the line radiation is too weak), the dashed lines turn over into dotted lines. The bold solid lines mark the mean values. The gray area indicates one standard deviation.}}
   \label{fig_correlations}
\end{figure*}

\begin{table}
\caption{\label{tabNNH}\footnotesize{Ratio of the N$_2$H$^+$ abundance in the core centers to the the maximum abundance, corresponding radius, hydrogen density, and dust temperature}}
\centering
\begin{tabular}{lcccc}
  \hline\hline
& Ratio & R$_\mathrm{max}$& $n_\mathrm{H,max}$& $T_\mathrm{D,max}$\\
& [\%] & [AU] &[cm$^{-3}$] & [K]\\
\hline
CB\,17 & 36 & 7500 & $4.7\times 10^4$ & 10.2\\
CB\,27 & 55 & 7700 & $6.5\times 10^4$& 11.7\\
B\,68 & 13 & 5400 & $1.4\times 10^5$& 9.5\\
CB\,244 & 41 & 6400 & $1.6\times 10^5$& 9.5\\
\hline
\end{tabular}
\end{table}

\begin{table}
\caption{\label{tabTime}\footnotesize{Chemical ages of the best fit models for $^{13}$CO, C$^{18}$O, and N$_2$H$^+$.}}
\centering
\begin{tabular}{lccc}
  \hline\hline
& Chem. Age& Chem. Age& Chem. Age \\
& of $^{13}$CO [yr] & of C$^{18}$O [yr]& of N$_2$H$^+$ [yr]\\
\hline
CB\,4 &... & $10^5$ & $10^5$\\
CB\,17 & $7\times 10^4$ & $4\times 10^4$ & $10^5$\\
CB\,26 & $4\times 10^5$ & $10^4$ & $7 \times 10^4$\\
CB\,27 & $2\times 10^5$ & $7\times 10^4$ & $10^5$\\
B\,68 &  $2\times 10^5$& $7\times 10^4$& $10^5$\\
CB\,244 &  $2\times 10^5$& $4\times 10^4$&$4\times 10^4$\\
\hline
\end{tabular}
\end{table}

In Fig.~\ref{fig_abundances}, we plot the modeled relative abundance
profiles of gaseous and frozen C$^{18}$O of the final models that correspond to the
emission profiles presented in Fig.~\ref{fig_mod_bestfits}. The
abundance of gaseous C$^{18}$O~($X$(C$^{18}$O$)\equiv n(\mathrm{C^{18}O})/n_\mathrm
H$) in molecular gas that is assumed to be shielded well and undepleted varies in the literature and ranges from $3.5\times
10^{-8}$~\citep{tafalla2004b} to $2.4\times
10^{-7}$~\citep{lee2003}. In the outer parts of all our cores, the C$^{18}$O~abundances $X$(C$^{18}$O) reach the value of \citet{tafalla2004b}. The highest abundance of $9\times 10^{-8}$ is reached in the envelope of the young and not that dense core CB\,26. The hydrogen densities in the regions of maximum C$^{18}$O~abundances are typically between $10^4$ and $5\times 10^4$\,cm$^{-3}$. In the center of the cores, the gaseous
C$^{18}$O abundance drops because of freezeout onto the dust grains. There, the C$^{18}$O abundance (relative to
the peak abundance) drops by a factor of 2 in the least dense core~(CB\,4) and by a factor of 30 in the denser cores~(CB\,27, B\,68, and
CB\,244). Towards the outer edges, CO is photodissociated by UV radiation, so that $X$(C$^{18}$O) there also drops. The emission profiles of $^{13}$CO are not reproduced as well by the chemical models as for C$^{18}$O~(see Fig.~\ref{fig_mod_bestfits}). The reason for this is most likely that the emission of this molecule is dominated from the outer envelope, which is probably warmer than the dust in the same layers. Thus, the observed emission profiles are flatter and extended further towards larger radii than is the case in the models. The optical thickness of the lines becomes apparent, for, instance, in the modeled emission profile of CB\,244~(Fig.~\ref{fig_mod_bestfits}). The density distribution of this model peaks at a radius of about $2\times 10^4$\,AU. The drop towards smaller radii is still indicated in the emission profile, but towards even smaller radii, it is dominated by the warm outer layer so that it even rises, thus explaining the double-peaked $^{13}$CO emission profiles in CB\,27 and CB\,244~(Fig.~\ref{fig_mod_bestfits}).

\begin{figure}
   \centering
   \includegraphics[width=0.5\textwidth]{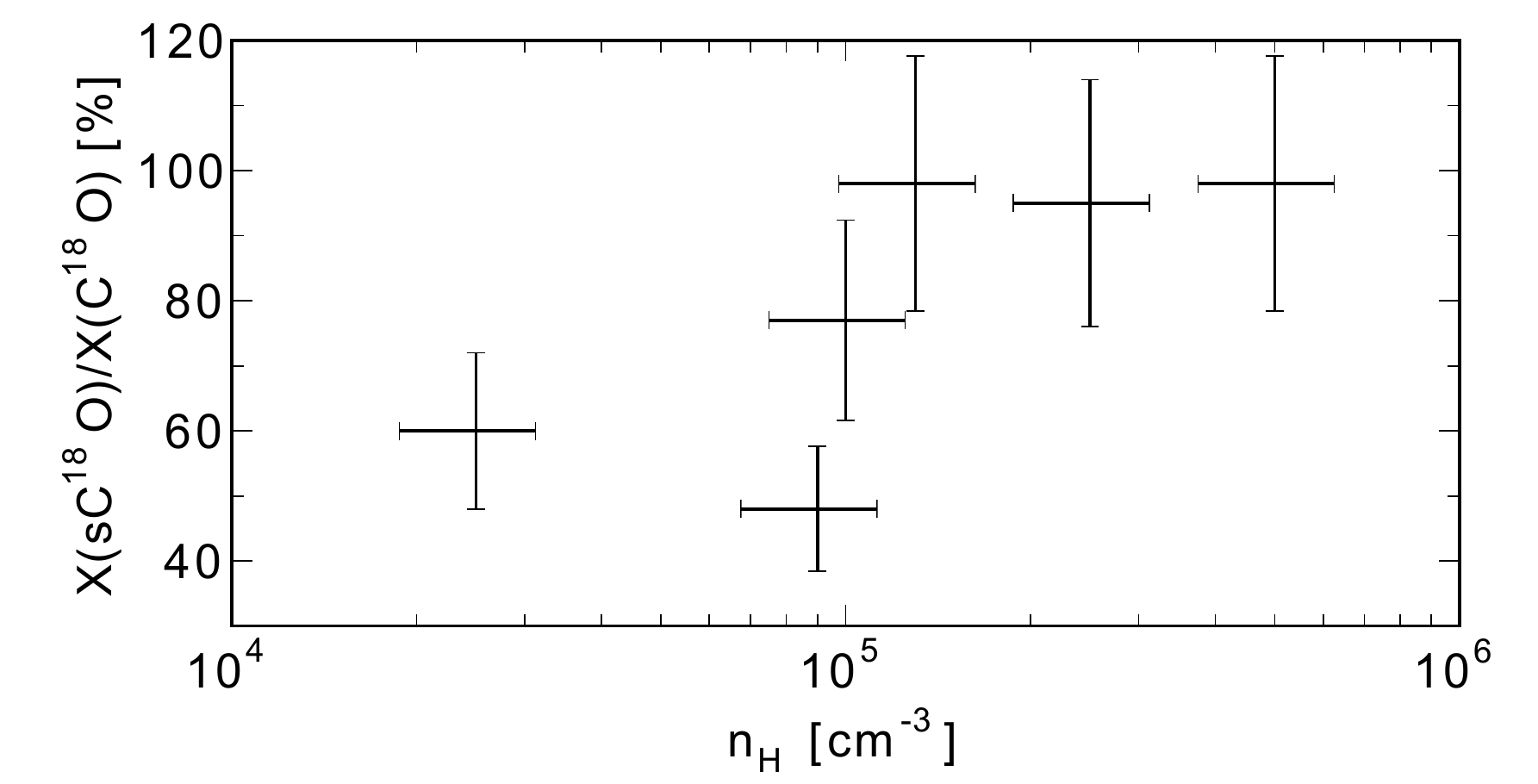}
   \caption{\footnotesize{Correlation of C$^{18}$O freezeout and hydrogen density in the core centers. The horizontal error bars on the hydrogen density represent the fitting uncertainty of the ray-tracing modeling (i.e., excluding systematic uncertainties from, e.g., the dust opacity model; see Sect.~\ref{sec_rt_ana}. The vertical error bars represent a 20\% uncertainty in the amount of icy CO in the chemical model.}}
   \label{fig_correlations_center}
\end{figure}

We also plot the profiles of icy C$^{18}$O since they are predicted by the best-fit chemical models
in Fig.~\ref{fig_abundances}. Its amount in the core center is in some
cases greater than the amount of gaseous C$^{18}$O at its peak
position, since at high densities the CO production is faster than the freezeout of atomic C. If the abundances
of icy C$^{18}$O at the core centers are compared to the total C$^{18}$O abundance there, the
ratios range from 45\% in CB\,4 to more than 90\% in those cores that have a central hydrogen density of more than $10^5$\, cm$^{-3}$. In Fig.~\ref{fig_correlations}, we compare these ratios in all cores and correlate them to the core radius, the hydrogen volume and column densities, and the dust temperature. The freezeout increases towards the core centers and with increasing hydrogen densities. The total temperature range between $T_\mathrm{in}$ and $T_\mathrm{out}$ in the cores is less than 10\,K, and the correlation of freezeout with the temperature is not as clear as for the other quantities. Typically, half of the total C$^{18}$O molecules stick to the grains at a radius of $\sim 10^4$\,AU, a hydrogen density of $\sim 10^5$\, cm$^{-3}$, and a temperature of 11\,K. In Table~\ref{tabCO} we list the maximum level of freezeout in the individual cores and the conditions at which half of the total C$^{18}$O molecules are frozen onto the dust grains. The correlation of the freezeout level to the hydrogen densities is plotted for the conditions of the core centers in Fig.~\ref{fig_correlations_center}.

\subsubsection{N$_\mathsf{2}$H$^+$ abundances}
\label{sec:chem_n2hp}

N$_2$H$^+$ forms and depletes on longer time scales than CO. We find that in four out of the six studied globules (those with a central density over $10^5$\, cm$^{-3}$), N$_2$H$^+$ is strongly depleted in the core centers. For comparison, we show in Appendix~\ref{sec:LTE} that a simple line-of-sight, averaged LTE-analysis would not reveal depletion of N$_2$H$^+$, although the emission is optically thin ($\tau \lesssim 0.5$) in all cores as we have derived for  calculating the molecular column densities in Appendix~\ref{sec:LTE}. The ratios of the central to maximum abundance of N$_2$H$^+$ molecules are listed in Table~\ref{tabNNH}. They range from 13\% to 55\%. The radii, hydrogen density, and dust temperature at which the N$_2$H$^+$ abundances reach their maximum are also given in Table~\ref{tabNNH}. CB\,4 and CB\,26 do not show detectable depletion of N$_2$H$^+$. CB\,4 is probably not dense enough and CB\,26 might be too young to show depletion at its low central density~(see Sect.~\ref{subsec:age}). The correlation of $X$(N$_2$H$^+$) with core radii, hydrogen volume, and column density and with temperature are plotted for all cores in the bottom row of Fig.~\ref{fig_correlations}. $X$(N$_2$H$^+$) typically peaks at hydrogen densities of $10^5$\,cm$^{-3}$ in cores at a chemical age of $10^5$\,yr. The mean radius of the peak $X$(N$_2$H$^+$) is ($6800\pm 1000$)\,AU. We resolve the regions of N$_2$H$^+$ depletion in all cores, since these radii are two to three times larger than the half beamwidth.

\begin{figure}
   \centering
   \includegraphics[width=0.5\textwidth]{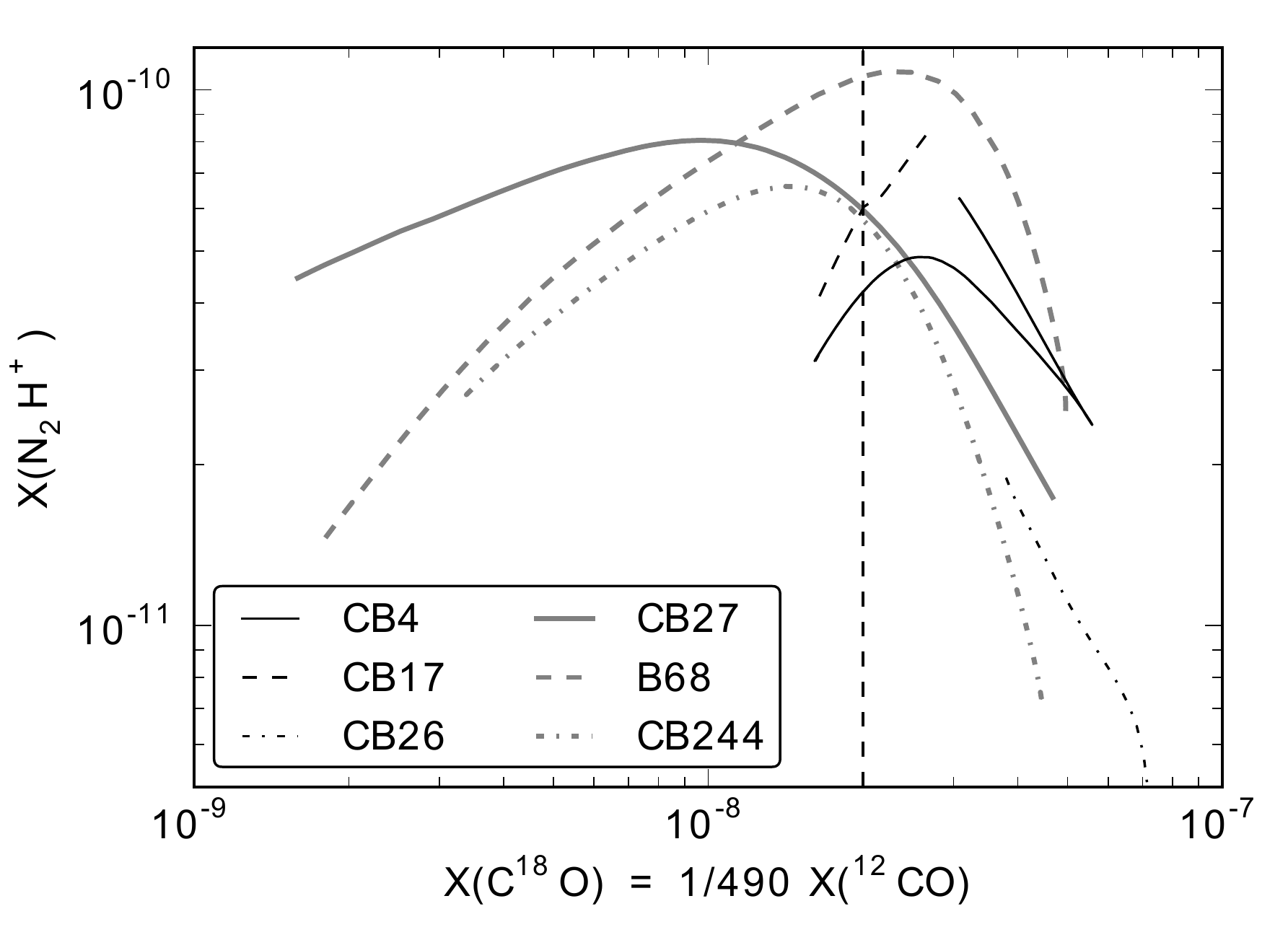}
   \caption{\footnotesize{Relative gas phase abundance of C$^{18}$O against that of N$_2$H$^+$. At $X$(C$^{18}$O$)\approx 2 \times 10^{-8}$ the correlations turns over into an anti-correlation. While at high $X$(C$^{18}$O), the N$_2$H$^+$ molecules are destroyed by reactions with CO, at low $X$(C$^{18}$O) (and high densities) freezeout of nitrogen is the dominant reason for depletion of N$_2$H$^+$.}}
   \label{fig_NNH_COcorrelations}
\end{figure}

We find that the depletion of N$_2$H$^+$ is more restricted to the center as compared to the depletion of CO. It occurs at hydrogen densities $\gtrsim 10^5$\,cm$^{-3}$ and column densities $>10^{22}$\,cm$^{-2}$, and requires a chemical age of $\gtrsim 10^5$\,yr. This finding is consistent with the work of \citet{bergin2002} on B\,68, while other studies have claimed that depletion of N-bearing species would only occur at hydrogen densities $\gtrsim 10^6$\,cm$^{-3}$~\citep{tafalla2002,pagani2005}. The abundance quickly increases outwards and peaks where the majority of CO is still frozen onto the grains. Farther out, it drops where CO becomes abundant in the gas phase. In Fig.~\ref{fig_NNH_COcorrelations}, we plot the relative gas phase C$^{18}$O abundance against that of N$_2$H$^+$ for all globules. The N$_2$H$^+$ and C$^{18}$O abundances are anticorrelated at C$^{18}$O abundances above $10^{-8}$ (corresponding to $X\mathrm{(CO)}=5\times 10^{-6}$ when assuming $X$(CO)/$X$(C$^{18}\mathrm{O})=490$), but the curves turn over to a positive gradient below this value. At this point, depletion of nitrogen mainly in form of N and N$_2$ becomes the dominant reason for depletion of N$_2$H$^+$ compared to reactions with CO.

\subsubsection{Chemical age of the cores}
\label{subsec:age}

"Chemical age" is a parameter in the modeling process. It sets the time span during which the gas develops starting from a state of purely atomic gas of 13 different elements. In this period complex molecules build up and deplete onto the grains. We find that the individual species are fitted best for models of different chemical ages. They are listed in Table~\ref{tabTime}. The average chemical age of the $^{13}$CO best-fit models is $(2\pm 1)\times
10^5$\,yr, for C$^{18}$O $(6\pm 3)\times
10^4$\,yr, and for N$_2$H$^+$ $(9\pm 2)\times
10^4$\,yr. Up to an age of $\sim 2\times 10^5$\,yr, the chemical models evolve quickly so that deviations of a few $10^4$\,yr already lead to significantly worse fits of the models to the data. Therefore, the difference in the chemical ages of the best-fit models between the individual species is significant. $^{13}$CO that mainly traces the outer envelope of the cores is modeled best with chemical ages that are on average more than $10^5$\,yr higher than those of the best-fit models of C$^{18}$O and N$_2$H$^+$. 

We interpret this finding as the result of core contraction and cooling during the chemical evolution. This would accelerate the freezeout of CO and the synthesis of N$_2$H$^+$ molecules. Since we assume a constant physical structure in the chemical modeling, we would thus systematically underestimate the chemical age of those species that trace the regions that have been contracting during the chemical evolution. Additional support for this hypothesis is provided by the analysis of the gravitational stability of the cores in Sect.~\ref{sec:stability}. We found there that four cores are thermally supercritical and might be gravitationally unstable. This finding is less profound for the other two cores, but cannot be excluded. 

Other effects that are not included in our model, such as nonthermal desorption processes like explosive desorption~\citep{shalabiea1994} or exothermic reactions in the grain mantles~\citep{garrod2007}, might also lead to underestimating the chemical ages. Also, turbulent diffusion~\citep{xie1995,willacy2002} of the gas and grains within the cores can decelerate the chemical aging. Coagulation of the dust grains in the dense part would also slow down the chemical evolution, since by this the grain surface area and thus the rate of grain surface reactions are reduced. In Sect.~\ref{sec:disc_grain} we show, however, that the bulk of the dust in these globules cannot be significantly processed and coagulated. Accounting for these combined intrinsic uncertainties in the density structures of the cores, uncertainties associated with chemical reaction rates, and a range of grain sizes in the depletion zone of the cloud, we estimate that the best-fit chemical ages are between $10^5$ and $10^6$~years. This is consistent with a prestellar core lifetime of 0.3-1.6~Myr as derived by~\citet{lee1999}.

\subsubsection{Strength of the impacting UV-field}
\label{sec:uv}

Changing the parameter $A_\mathrm{v,\,out}$ that accounts for the unconstrained extent of the dusty material around the globules in the chemical models has
effects that cannot always be clearly distinguished from changing
either the hydrogen density or the gas temperature. Nevertheless, the observations are modeled better if $A_\mathrm{v,\,out}$ is increased for half of the globules by a few mag relative to the values that have been
measured with NIR extinction mapping~(which range from 0.2 to 2\,mag
for the different globules at the model boundaries). This is probably not a result  of overestimating the ISRF in the chemical models. \citet{launhardt2013} derived the total FIR/mm luminosity of the globules, and it is approximately equal for all of
them to the luminosity of the "standard UV-field"~\citep{draine1978} impacting on spheres of the size of the globules. The same field is also adopted in the chemical modeling, so we do not overestimate its strength.

In the envelopes of the globules at hydrogen densities below $10^4$\,cm$^{-3}$, gas and dust temperatures are expected to be decoupled. Since we assume equal gas and dust temperatures in the modeling, we introduce systematic errors in our analysis. In particular, the emission of the gas in the envelopes is underestimated. With a large $A_\mathrm{v,\,out}$ we artificially compensate for this error. The additional shielding protects the molecules from being photodissociated and thus increases their abundance. This hypothesis is supported when comparing $A_\mathrm{v,\,out}$ of the best models to $A_\mathrm{v,\,NIR}$, which are the values derived from the NIR extinction maps. Both values appear to be anticorrelated, and while we do not need to increase $A_\mathrm{v,\,out}$ with respect to the observed values at $A_\mathrm{v,\,NIR} \sim 2$\,mag, we have to increase it by more than 3~mag for the least shielded cores. Because of these uncertainties, we think that the need to increase $A_\mathrm{v,\,out}$ is not convincing evidence that the globules are in fact shielded more strongly from UV-radiation than inferred from NIR-extinction maps. 

\begin{figure}
   \centering
   \includegraphics[width=0.5\textwidth]{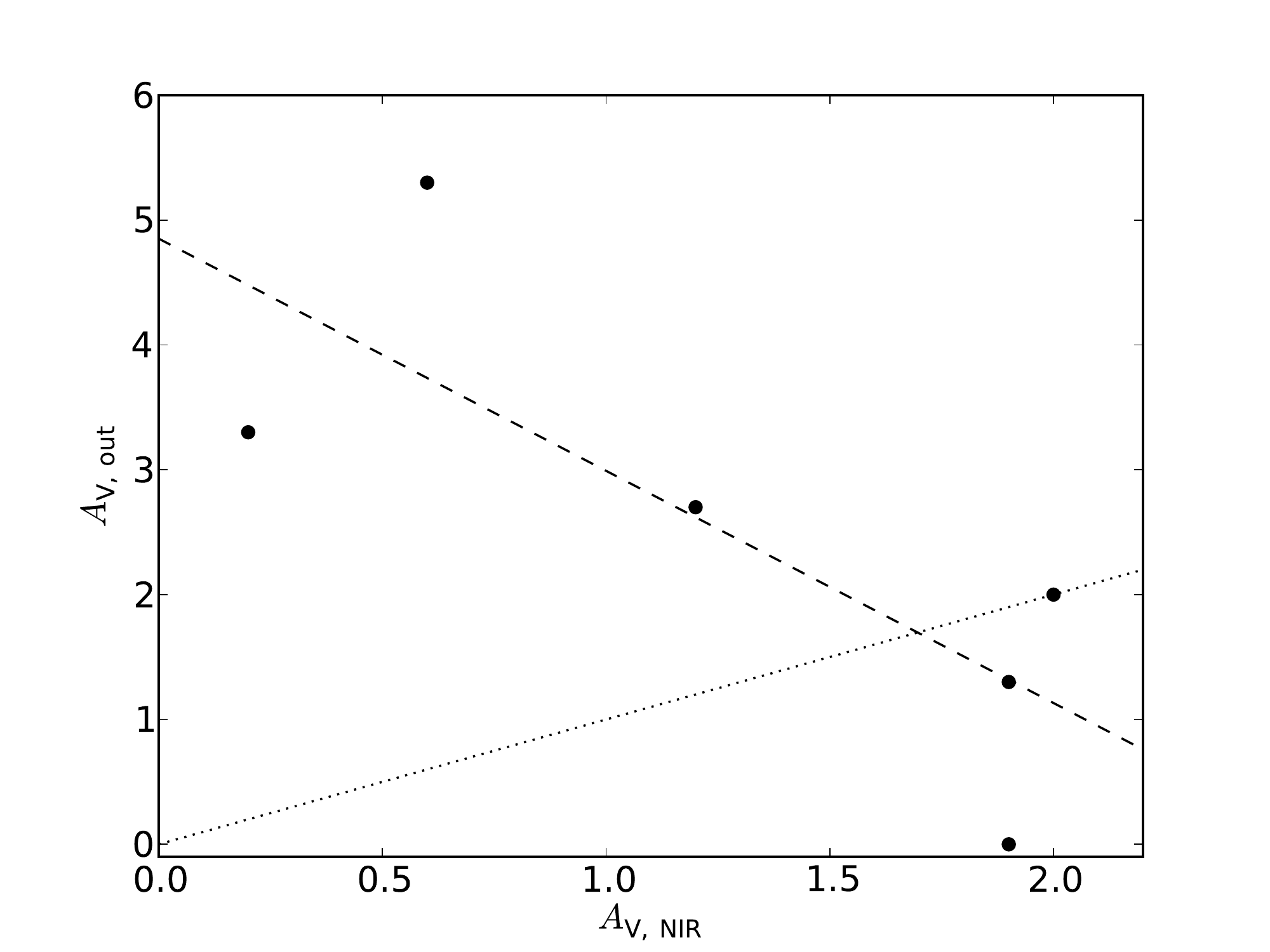}
   \caption{\footnotesize{We plot the parameters $A_\mathrm{V,\,out}$ of the best models over the measured values of $A_\mathrm{V,\,NIR}$ from the NIR extinction maps. The dashed line is a linear fit to the data points, which is only meant to illustrate the trend. The dotted line indicates $A_\mathrm{V,\,out}=A_\mathrm{V,\,NIR}$. We interpret the anti-correlation as indication of increasing decoupling of gas and dust temperatures towards lower extinction.}}
   \label{fig_av_av}
\end{figure}

A direct measurement of the gas temperature in the envelopes would of course be desirable as a constraint for the chemical modeling. Furthermore, in comparison to the dust temperature it would allow distinguishing between shielding from extended envelopes and a weaker ISRF as the presumed Draine field, since the gas is only heated from the FUV component of the ISRF. Measuring the excitation of the gas in the envelopes is difficult, however, since the emission is weak, and most transitions are subthermally excited. The best-suited transition to study for this purpose might therefore be the $^{12}$CO~(J=1-0) line. Unfortunately, we do not have such data at hand. Another way to distinguish between both scenarios may be to study the distribution and emission of PAHs. All globules of this study have been mapped with \textit{Spitzer's} infrared array camera (IRAC). In particular the 5.8\,$\mu$m and 8\,$\mu$m bands are known to show PAH emission~\citep[e.g.,][]{pagani2010}. However, the maps are not extended enough to define a background level that would allow absolute values of the emission to
be derived.

\subsubsection{Constraints on the grain properties}
\label{sec:disc_grain}

To correctly model the observed molecular abundances, we must increase the molecular hydrogen density and, with it, the initial abundances of all species by a factor of 2 to 3 (except for B\,68). We thus seem to systematically underestimate the hydrogen density in the dust-based modeling. This is understandable, at least in the envelopes of the cores, since we have assumed mildly coagulated grains, with a coagulation time of $10^5$\,yr at a gas density of $10^5$\,cm$^{-3}$~\citep{oh94}\footnote{ftp://cdsarc.u-strasbg.fr/pub/cats/J/A+A/291/943} over the entire globule. With the assumption of normal ISM dust~(e.g., OH1), we would have derived hydrogen densities that are higher by a factor of 2.5. Increasing the density could also partly compensate for an underestimation of the gas temperature. However, we do not find any dependence on the observed $A_V$. Moreover, the only core for which we do not have to increase the density is that of B\,68 -- one of the least shielded cores in our sample. Our molecular line observations and modeling thus suggest that in most regions of the cores that are resolved by our study, the grains are not yet significantly coagulated.

\section{Conclusions}

We observed a sample of seven globules containing starless cores in the $^{12}$CO~(J=2-1), the
$^{13}$CO~(J=2-1), the C$^{18}$O~(J=2-1) and the
$\mathrm{N_2H^+}$~(J=1-0) transitions. The globules were also observed
with the Herschel bolometers as part of the EPoS project~\citep{launhardt2013}. We presented dust
temperature and hydrogen density maps of the globules that were obtained
with a ray-tracing technique. Based on these dust-based temperature and density profiles, we analyzed the molecular
emission assuming $T_\mathrm{Dust}=T_\mathrm{Gas}$. In the following we summarize the main results of this study.

\begin{enumerate}

\item We confirm the findings of \citet{launhardt2013} on the stability of the cores. Only one starless core~\citep[CB\,244, see also][]{stutz2010} is clearly not supported against gravitational collapse by thermal pressure. CB\,17, CB\,27, and B\,68 are probably also thermally supercritical. CB\,4 and CB\,26 are probably gravitationally stable.

\item CO is depleted in the center of all studied cores to a level of at least 46\%. CB\,27, B\,68, and CB\,244 show a central CO depletion of more than 90\%. The level of freezeout increases towards the core centers and with the hydrogen density. The average radius at which half of the CO is frozen to the grains is 9000\,AU, and the average hydrogen density is $10^5$\,cm$^{-3}$.

\item $\mathrm{N_2H^+}$ is depleted in the center of the cores with hydrogen densities exceeding $10^5$\,cm$^{-3}$. The radius with maximum abundance of $\mathrm{N_2H^+}$ is on average at 6800\,AU. The degree of depletion in the center with respect to the maximum abundance ranges from 13\% to 55\%.

\item The chemical age at which we find that the models match the data best is higher for $^{13}$CO than for C$^{18}$O and $\mathrm{N_2H^+}$. We find an average age of the six modeled cores of $2\pm 1\times 10^5$~yr for $^{13}$CO and of $6\pm 3\times 10^4$~yr for C$^{18}$O, and  $9\pm 2\times 10^4$~yr for $\mathrm{N_2H^+}$.  The different ages of the dense gas tracers and $^{13}$CO suggest a central contraction of the cores during the chemical evolution. Considering the uncertainties, all cores have a chemical age of at most $10^6$\,yr and are probably younger.

\item We generally need to increase the gas density that has been derived from the dust density, assuming mildly coagulated grains$^8$ by a factor of 2-3 in order to model the observed molecular emission profiles correctly. We interpret this as a sign that the dust grains are not as heavily coagulated as assumed by our dust model in large parts of the globules and are better described by ISM dust models like OH1.

\item We have to reduce the strength of the UV radiation field impacting at the globule boundaries with respect to the value predicted by \citet{draine1978} for half of the cores. However, we think that this does not indicate that the impacting UV-field is in fact weaker, but rather that we thereby compensate for the systematic error that is introduced by the underestimation of the gas temperature in the envelopes of these cores.

\end{enumerate}

\begin{acknowledgements}
 We thank the anonymous referee and the editor M. Walmsley for comments and suggestions that helped to improve
the clarity and completeness of the paper. We wish to thank the IRAM Granada staff for the support at the 30m telescope and T.~Bergin for providing us with his line data of B\,68. We also thank N.~J.~Evans~II and E.~Keto for interesting discussions that helped to improve the clarity of the paper. PACS has been developed by a consortium of institutes led by MPE (Germany), including UVIE (Austria); KU Leuven, CSL, IMEC (Belgium); CEA, LAM (France); MPIA (Germany); INAF-IFSI/OAA/OAP/OAT, LENS, SISSA (Italy); IAC (Spain). This development has been supported by the funding agencies BMVIT (Austria), ESA-PRODEX (Belgium), CEA/CNES (France), DLR (Germany), ASI/INAF (Italy), and CICYT/MCYT (Spain). The Heinrich Hertz Submillimeter Telescope is operated by Steward Observatory at the University of Arizona with partial support from the U.S. National Science Foundation through grant AST-1140030.
D.S. acknowledges support by
the Deutsche Forschungsgemeinschaft through SPP 1385: "The first ten million
years of the solar system -- a planetary materials approach" (SE 1962/1-1 and
SE 1962/1-2). The work of A.M.S. was supported by the Deutsche
Forschungsgemeinschaft priority program 1573 ("Physics of the
Interstellar Medium"). Y.P. was supported by the Russian Foundation
for Basic Research (project 13-02-00642). H.L., M.N., and Z.B. were funded by the Deutsches Zentrum f\"ur Luft- und Raumfahrt (DLR).

\end{acknowledgements}

\bibliographystyle{aa}
\bibliography{test}

\begin{appendix}

\section{Observations}


\begin{figure*}
   \centering
   \includegraphics[width=0.95\textwidth]{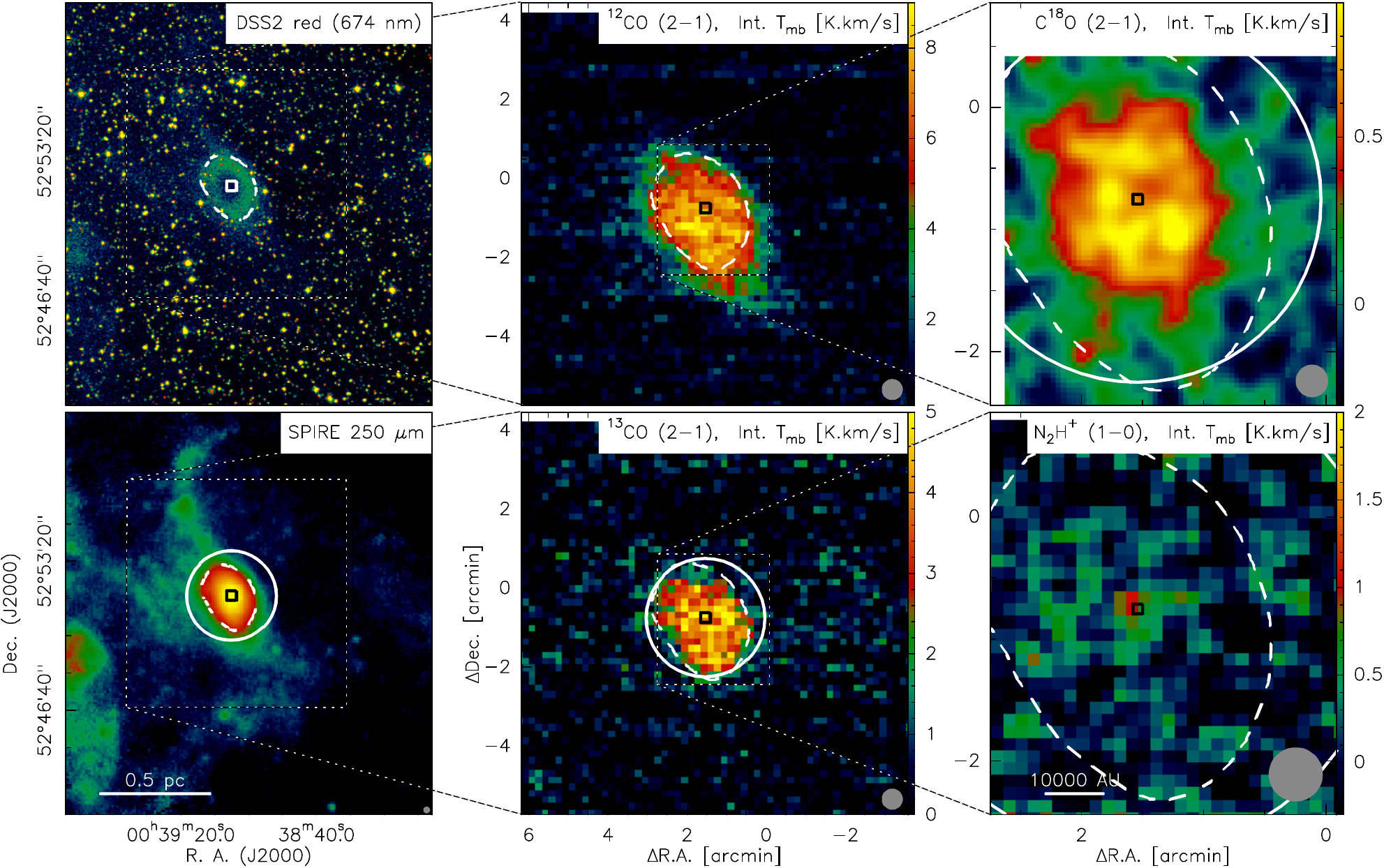}
   \caption{\footnotesize{Observations of CB4. Digitized Sky Survey (DSS) red, Herschel SPIRE 250~$\mu$m, $^{12}$CO~(J=2-1), $^{13}$CO~(J=2-1), C$^{18}$O~(J=2-1), and N$_2$H$^+$~(J=1-0). The gray circles in the lower right corners indicate the respective beam sizes. The squared marThese obserservations are discuker indicates the center of the
     starless core. The dashed contour marks N$_{\rm H}=10^{21}$cm$^{-2}$. The white circles indicate the regions of which 1D-profiles were obtained by azimuthally averaging.}}
   \label{fig_obs_cb4}
\end{figure*}

\begin{figure*}[b]
   \centering
   \includegraphics[width=0.95\textwidth]{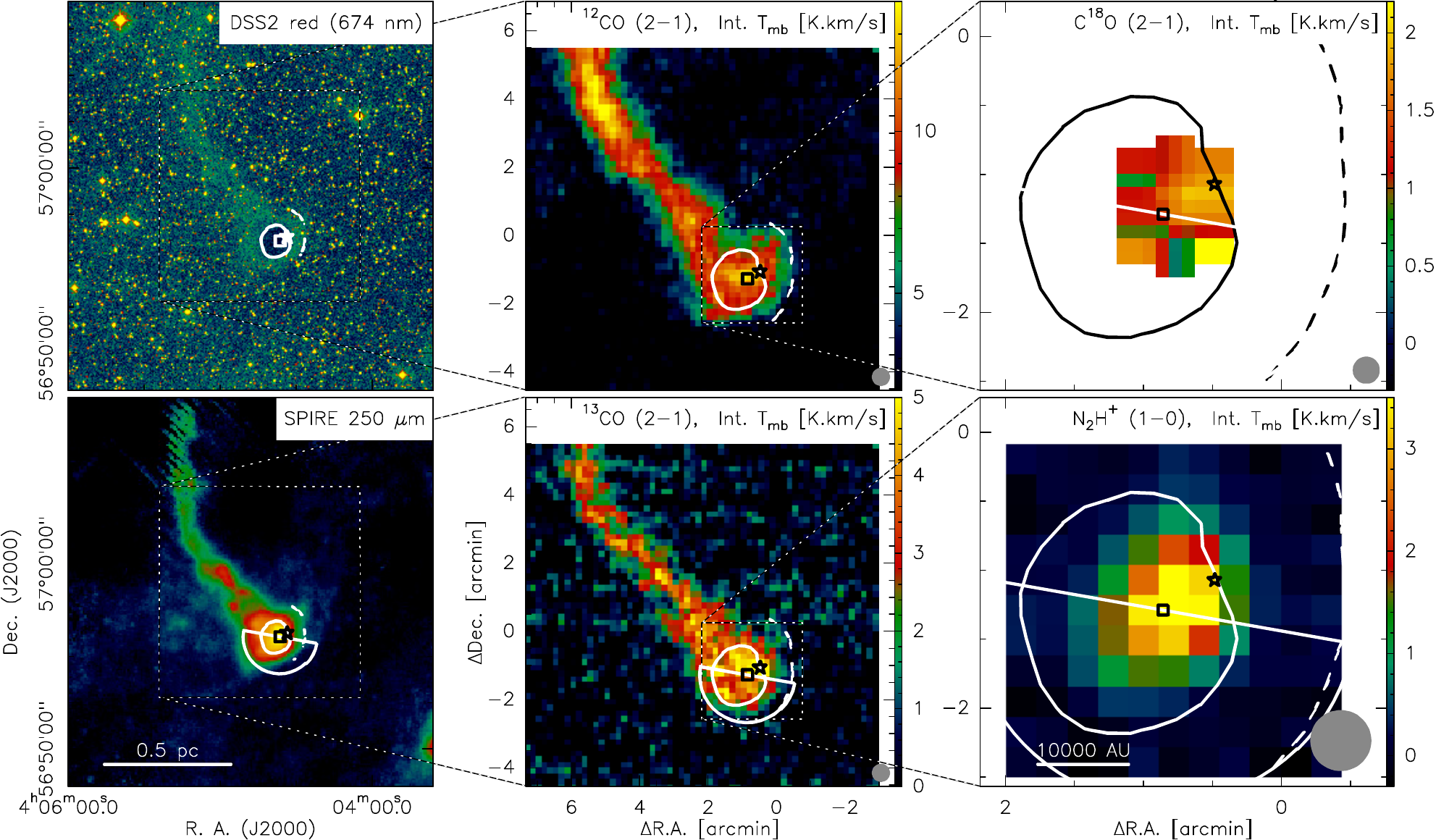}
   \caption{\footnotesize{Observations of CB17. DSS red, Herschel SPIRE 250~$\mu$m, $^{12}$CO~(J=2-1), $^{13}$CO~(J=2-1), C$^{18}$O~(J=2-1), and N$_2$H$^+$~(J=1-0). The gray circles in the lower right corners indicate the respective beam sizes. The squared marker indicates the center of the
     starless core. The asterisk the position of the Class~I~YSO. The dashed contour marks N$_{\rm H}=10^{21}$cm$^{-2}$, the solid contour N$_{\rm H}=10^{22}$cm$^{-2}$. The white circles indicate the regions where 1D-profiles were obtained by azimuthally averaging.}}
   \label{fig_obs_cb26}
\end{figure*}

\begin{figure*}
   \centering
   \includegraphics[width=0.95\textwidth]{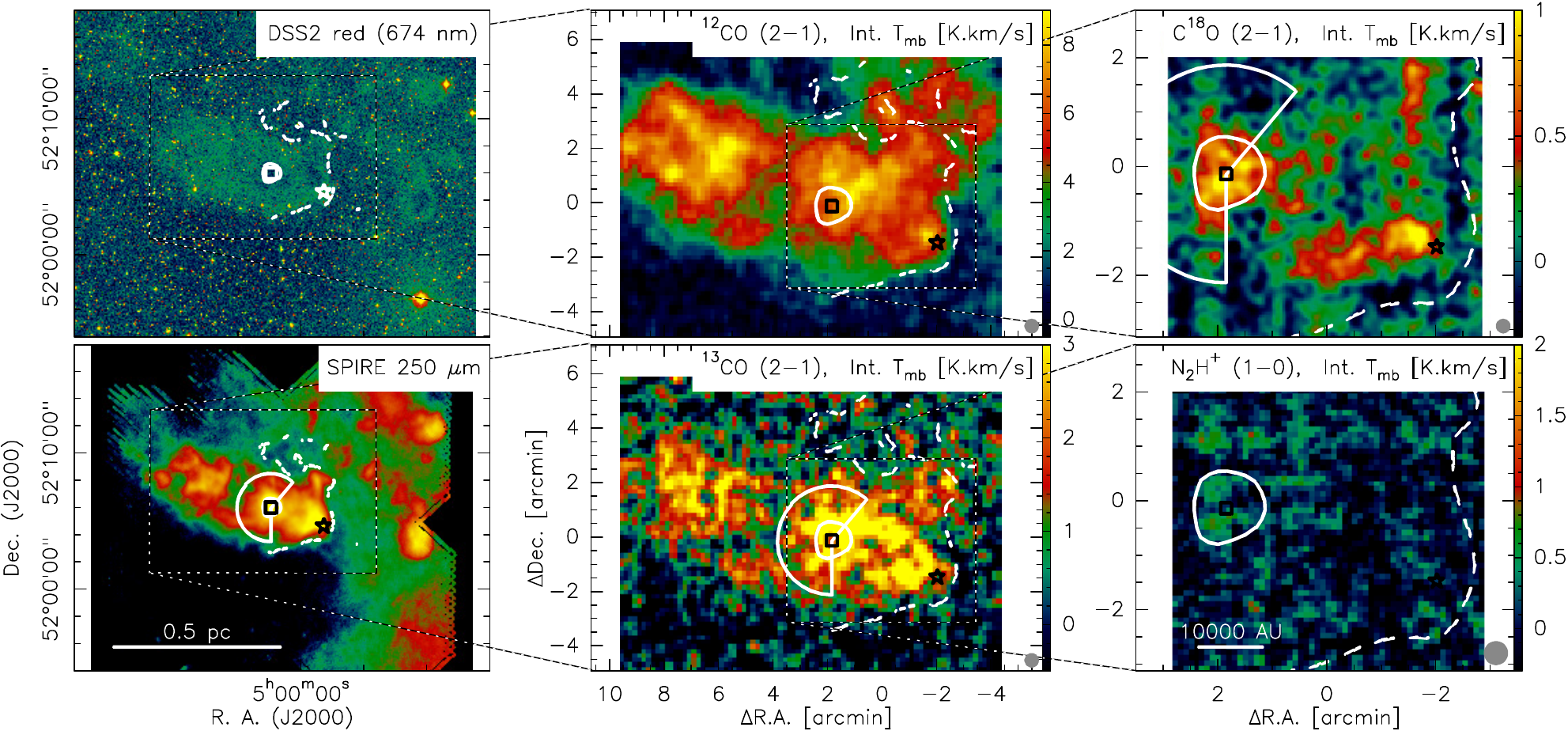}
   \caption{\footnotesize{Observations of CB26. DSS red, Herschel SPIRE 250~$\mu$m, $^{12}$CO~(J=2-1), $^{13}$CO~(J=2-1), C$^{18}$O~(J=2-1), and N$_2$H$^+$~(J=1-0). The gray circles in the lower right corners indicate the respective beam sizes. The squared marker indicates the center of the
     starless core, the asterisk the position of the Class~I~YSO. The dashed contour marks N$_{\rm H}=10^{21}$cm$^{-2}$, the solid contour N$_{\rm H}=10^{22}$cm$^{-2}$. The white circles indicate the regions where 1D-profiles were obtained by azimuthally averaging.}}
   \label{fig_obs_cb26}
\end{figure*}

\begin{figure*}
   \centering
   \includegraphics[width=0.95\textwidth]{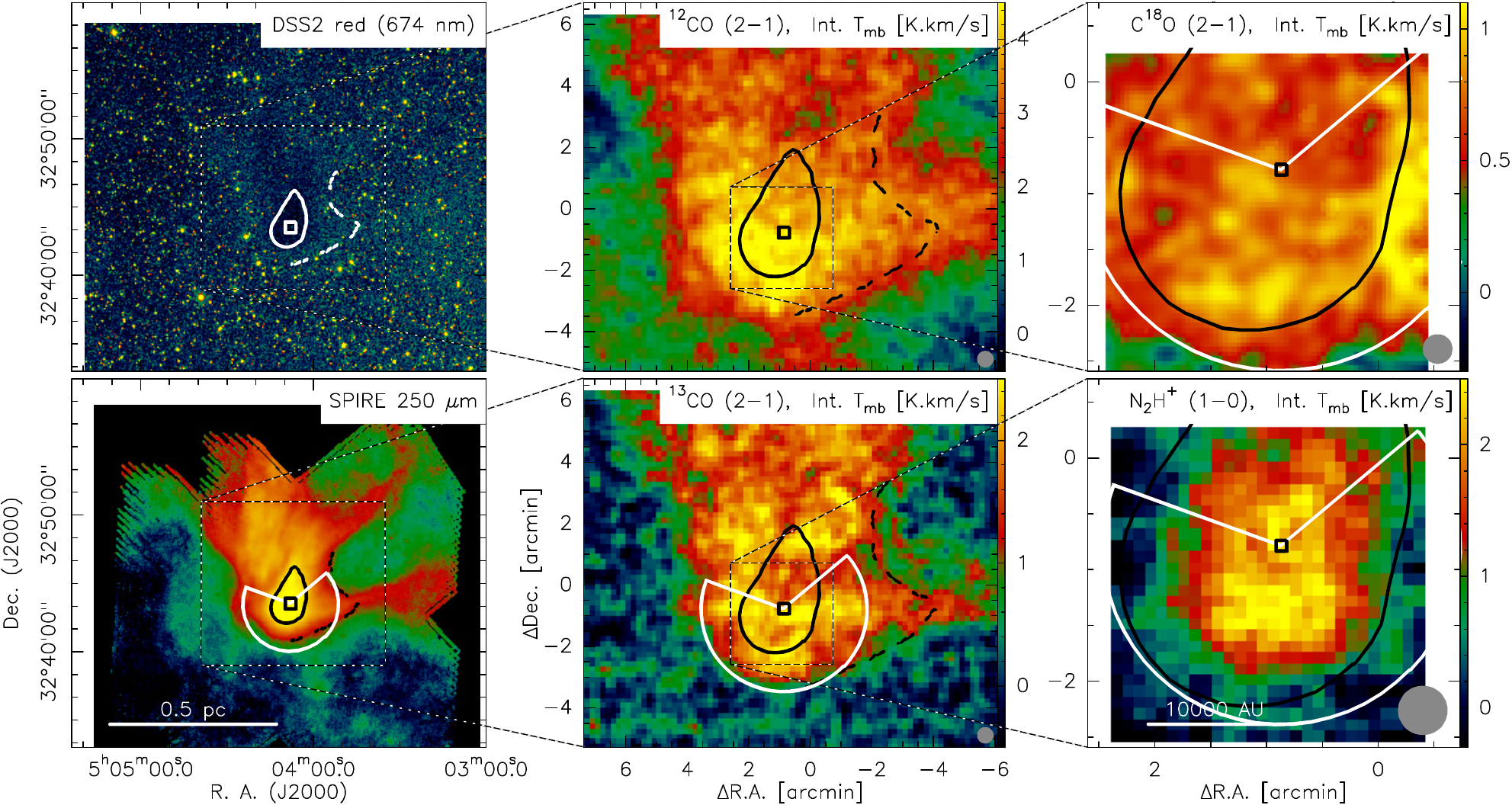}
   \caption{\footnotesize{Observations of CB27. DSS red, Herschel SPIRE 250~$\mu$m, $^{12}$CO~(J=2-1), $^{13}$CO~(J=2-1), C$^{18}$O~(J=2-1), and N$_2$H$^+$~(J=1-0). The gray circles in the lower right corners indicate the respective beam sizes. The squared marker indicates the center of the
     starless core. The dashed contour marks N$_{\rm H}=10^{21}$cm$^{-2}$, the solid contour N$_{\rm H}=10^{22}$cm$^{-2}$. The white circles indicate the regions where 1D-profiles were obtained by azimuthally averaging.}}
   \label{fig_obs_cb27}
\end{figure*}

\begin{figure*}[p]
   \centering
   \includegraphics[width=0.95\textwidth]{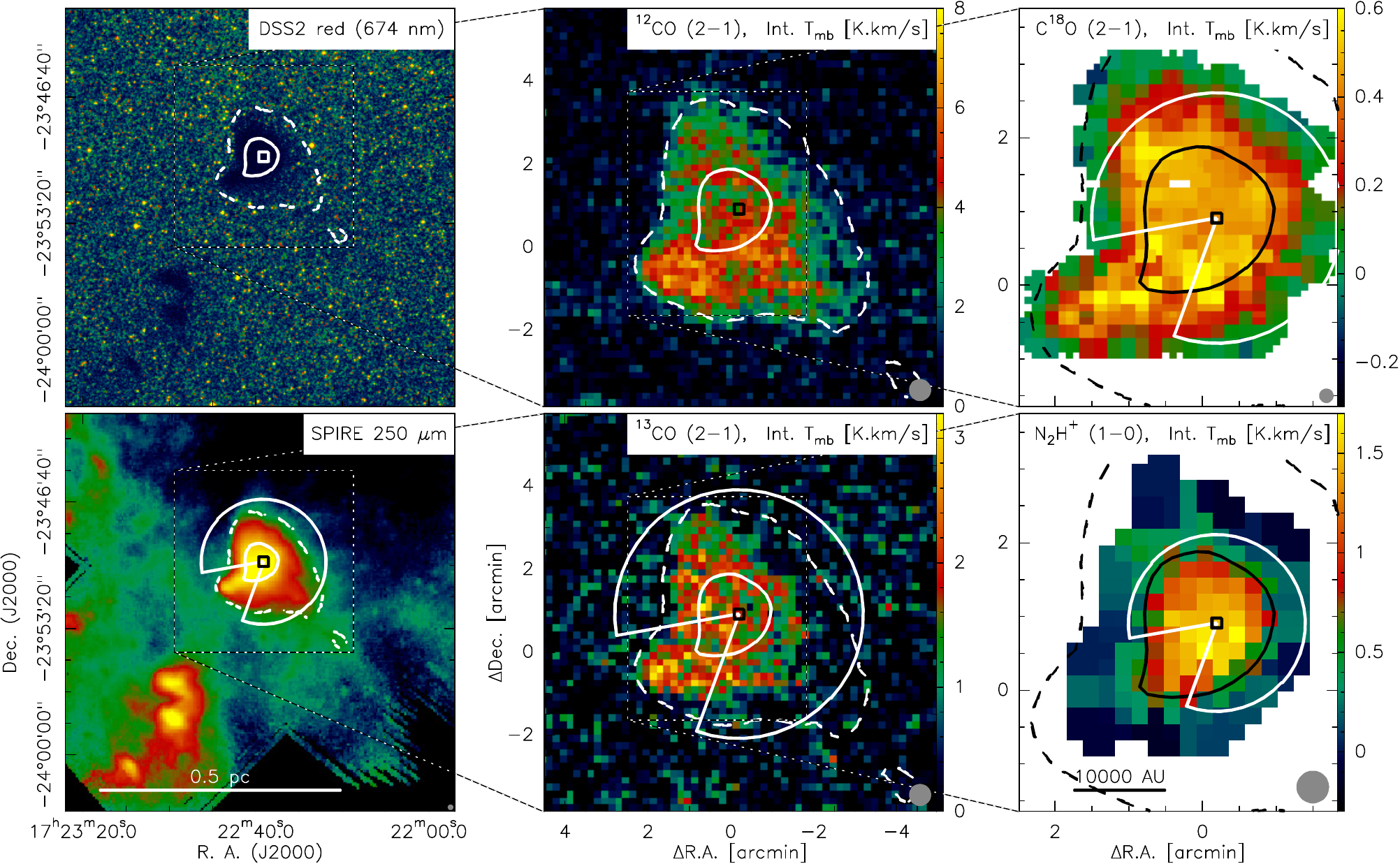}
   \caption{\footnotesize{Observations of B68. DSS red, Herschel SPIRE 250~$\mu$m, $^{12}$CO~(J=2-1), $^{13}$CO~(J=2-1), C$^{18}$O~(J=2-1), and N$_2$H$^+$~(J=1-0). The gray circles in the lower right corners indicate the respective beam sizes. The squared marker indicates the center of the
     starless core. The dashed contour marks N$_{\rm H}=10^{21}$cm$^{-2}$, the solid contour N$_{\rm H}=10^{22}$cm$^{-2}$. The white circles indicate the regions where 1D-profiles were obtained by azimuthally averaging.}}
   \label{fig_obs_b68}
\end{figure*}

\begin{figure*}[p]
   \centering
   \includegraphics[width=0.95\textwidth]{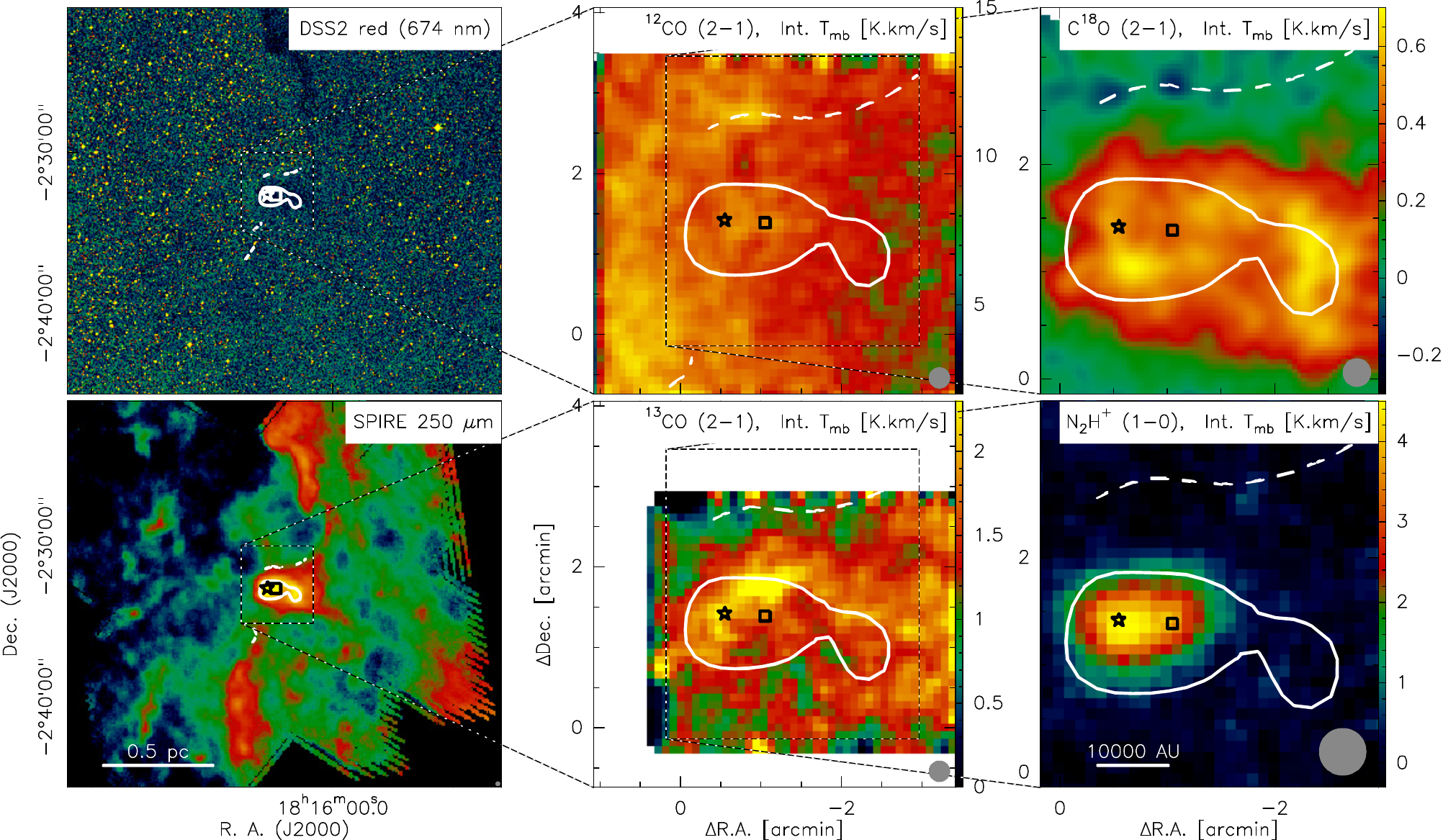}
   \caption{\footnotesize{Observations of CB130: DSS red, Herschel SPIRE 250~$\mu$m, $^{12}$CO~(J=2-1), $^{13}$CO~(J=2-1), C$^{18}$O~(J=2-1), and N$_2$H$^+$~(J=1-0). The gray circles in the lower right corners indicate the respective beam sizes. The squared marker indicates the center of the
     starless core. The asterisk the position of the Class~I~YSO. The dashed contour marks N$_{\rm H}=10^{21}$cm$^{-2}$, the solid contour N$_{\rm H}=10^{22}$cm$^{-2}$. The white circles indicate the regions where 1D-profiles were obtained by azimuthally averaging.}}
   \label{fig_obs_cb130}
\end{figure*}

\begin{figure*}[p]
   \centering
   \includegraphics[width=0.95\textwidth]{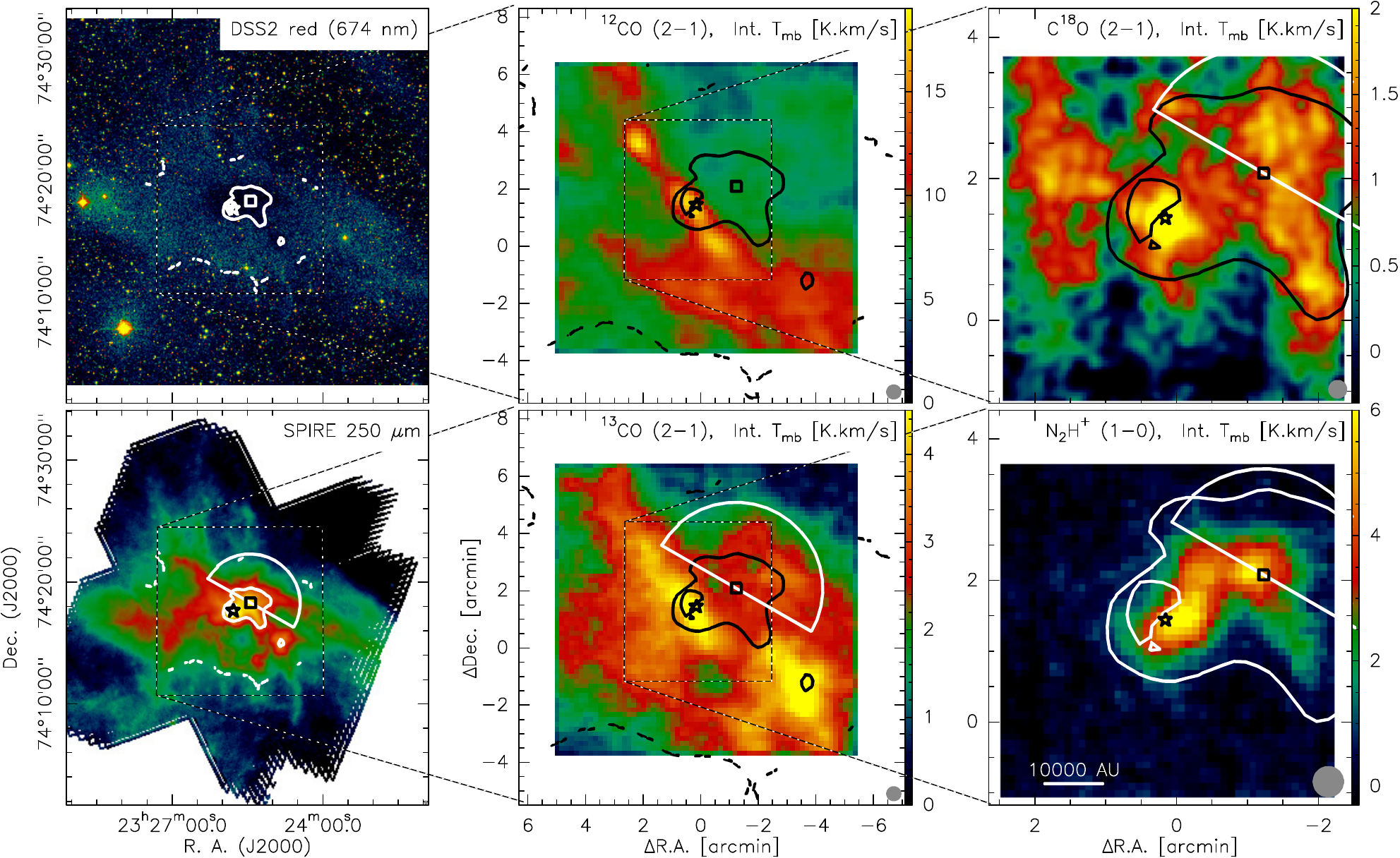}
   \caption{\footnotesize{Observations of CB244: DSS red, Herschel SPIRE 250~$\mu$m, $^{12}$CO~(J=2-1), $^{13}$CO~(J=2-1), C$^{18}$O~(J=2-1), and N$_2$H$^+$~(J=1-0). The gray circles in the lower right corners indicate the respective beam sizes. The squared marker indicates the center of the
     starless core. The asterisk in the center of the Class~0~protostellar core. The dashed contour marks N$_{\rm H}=10^{21}$cm$^{-2}$, the solid contour N$_{\rm H}=10^{22}$cm$^{-2}$. The white circles indicate the regions where 1D-profiles were obtained by azimuthally averaging.}}
   \label{fig_obs_cb244}
\end{figure*}

\pagebreak[4]

\afterpage{\clearpage
\section{Hydrogen density and dust temperature maps}


\begin{figure}[htpb]
   \centering
   \includegraphics[width=0.45\textwidth]{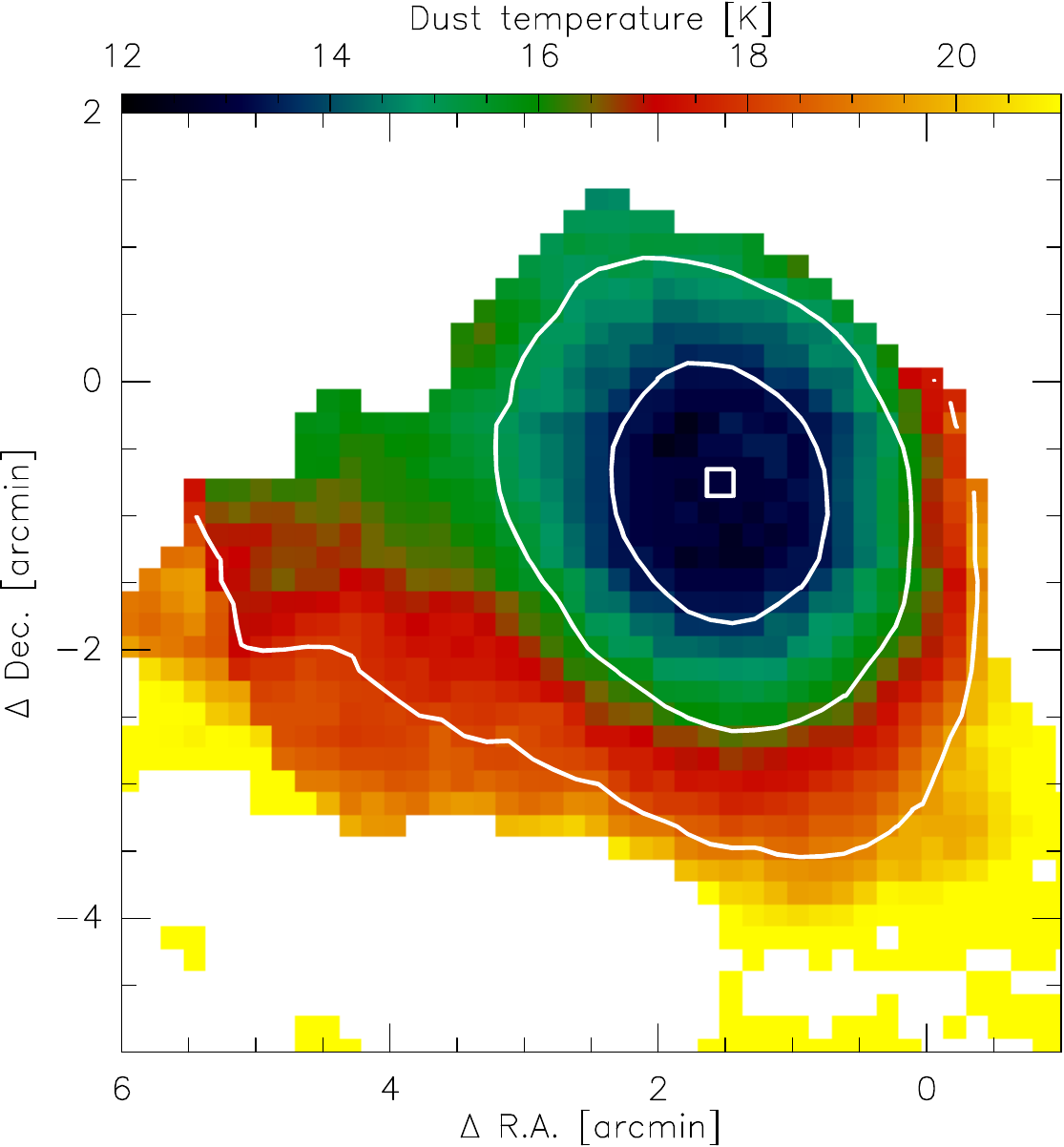}
   \caption{\footnotesize{Dust temperature map of CB4 overlaid with contours of the hydrogen density. They mark densities of $10^2$, $10^3$, and $10^4$~cm$^{-3}$. The square indicates the center of the
     starless core.}}
   \label{fig_dustmaps_CB4}
\end{figure}

\begin{figure}[htpb]
   \centering
   \includegraphics[width=0.45\textwidth]{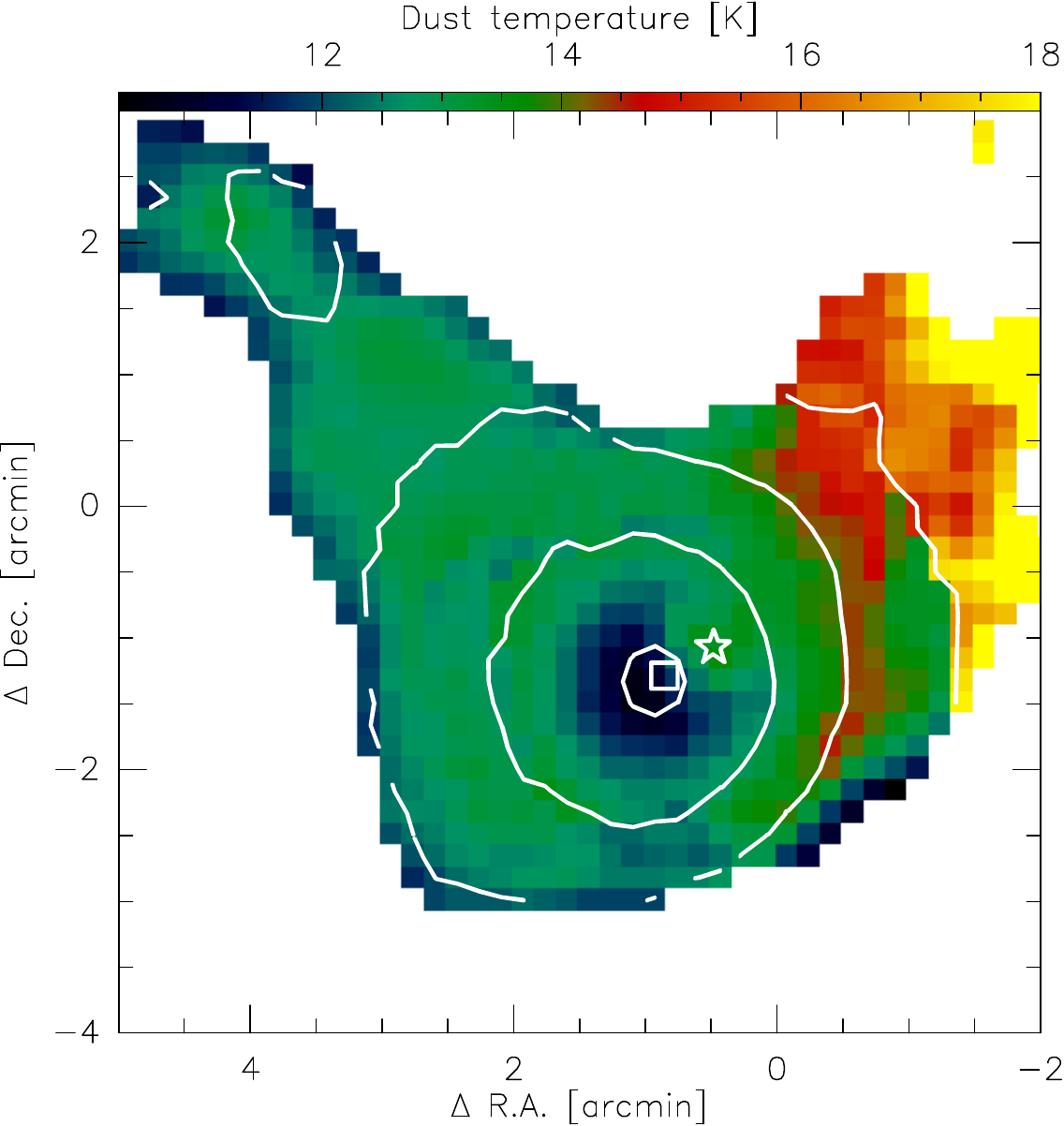}
   \caption{\footnotesize{Dust temperature map of CB17 overlaid with contours of the hydrogen density. They mark densities of $10^2$, $10^3$, $10^4$, and $10^5$~cm$^{-3}$. The square indicates the center of the
     starless core.}}
   \label{fig_dustmaps_CB17}
\end{figure}

\begin{figure}[htpb]
   \centering
   \includegraphics[width=0.45\textwidth]{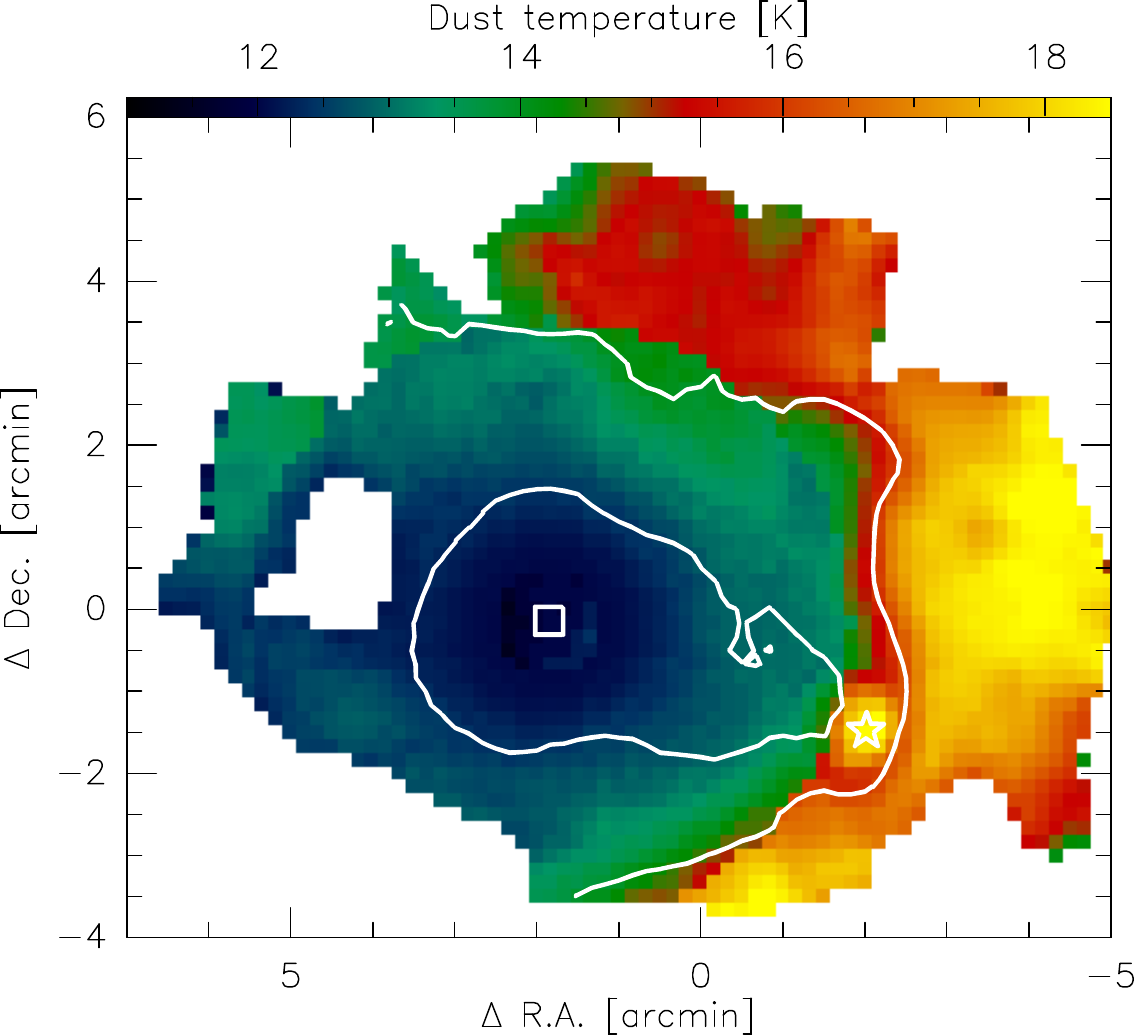}
   \caption{\footnotesize{Dust temperature map of CB26 overlaid with contours of the hydrogen density. They mark densities of $10^3$, and $10^4$~cm$^{-3}$. The square indicates the center of the
     starless core.}}
   \label{fig_dustmaps_CB26}
\end{figure}

\begin{figure}[htpb]
   \centering
   \includegraphics[width=0.45\textwidth]{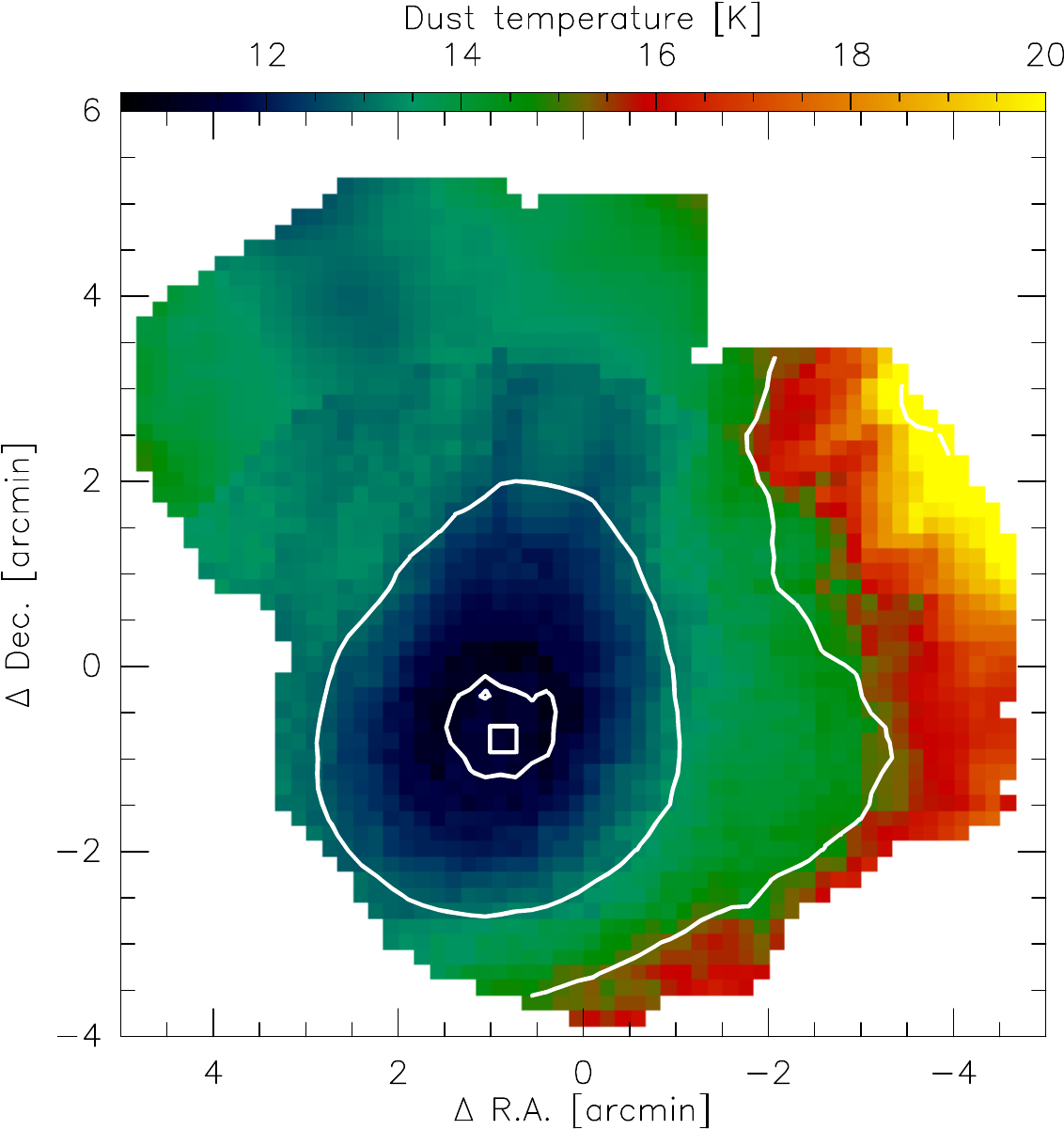}
  \caption{\footnotesize{Dust temperature map of CB27 overlaid with contours of the hydrogen density. They mark densities of $10^3$, $10^4$, and $10^5$~cm$^{-3}$. The square indicates the center of the
     starless core.}}
   \label{fig_dustmaps_CB27}
\end{figure}

\begin{figure}[htpb]
   \centering
   \includegraphics[width=0.45\textwidth]{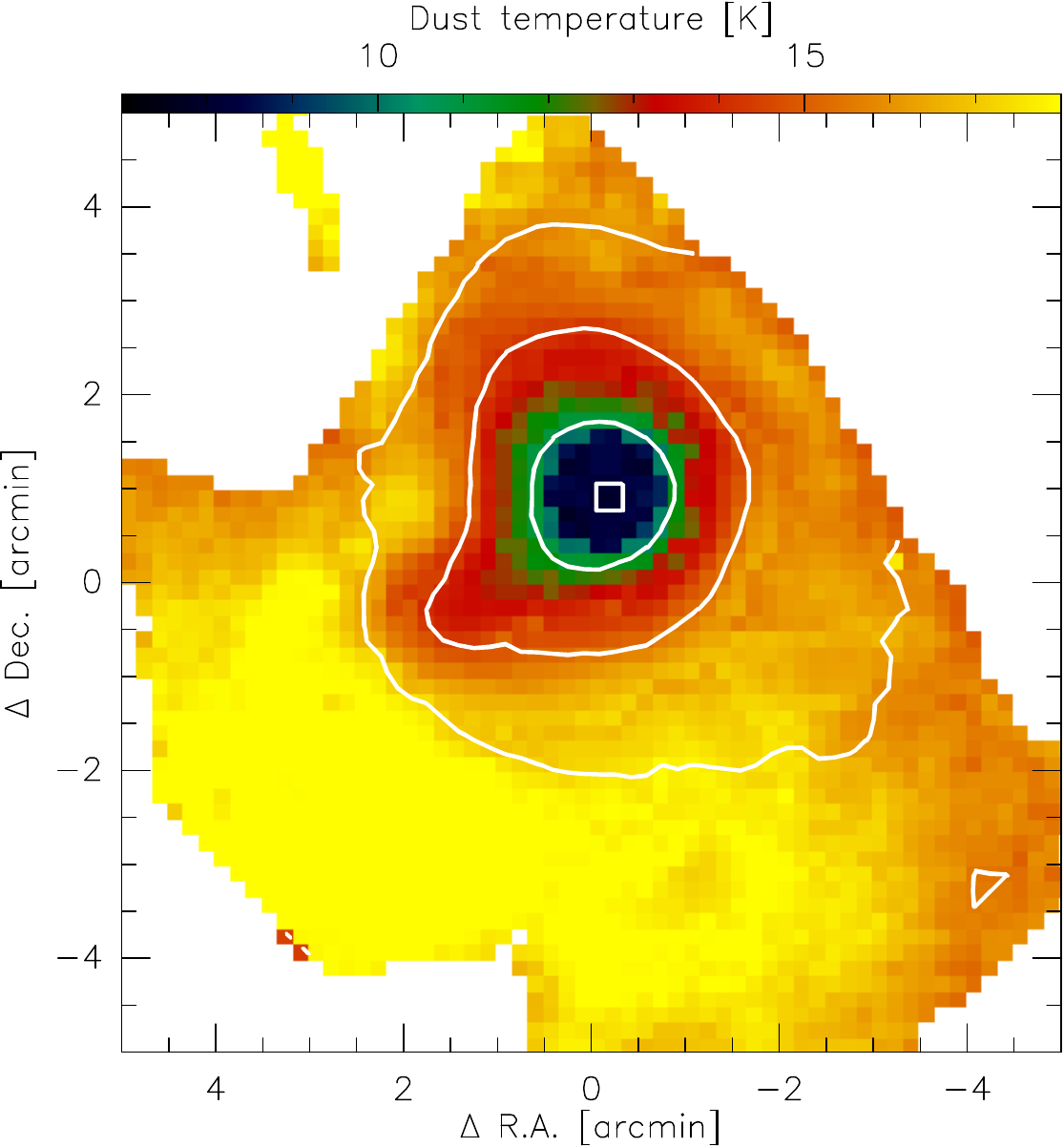}
   \caption{\footnotesize{Dust temperature map of B68 overlaid with contours of the hydrogen density. They mark densities of $10^3$, $10^4$, and $10^5$~cm$^{-3}$. The square indicates the center of the
     starless core.}}
   \label{fig_dustmaps_B68}
\end{figure}

\begin{figure}[htpb]
   \centering
   \includegraphics[width=0.45\textwidth]{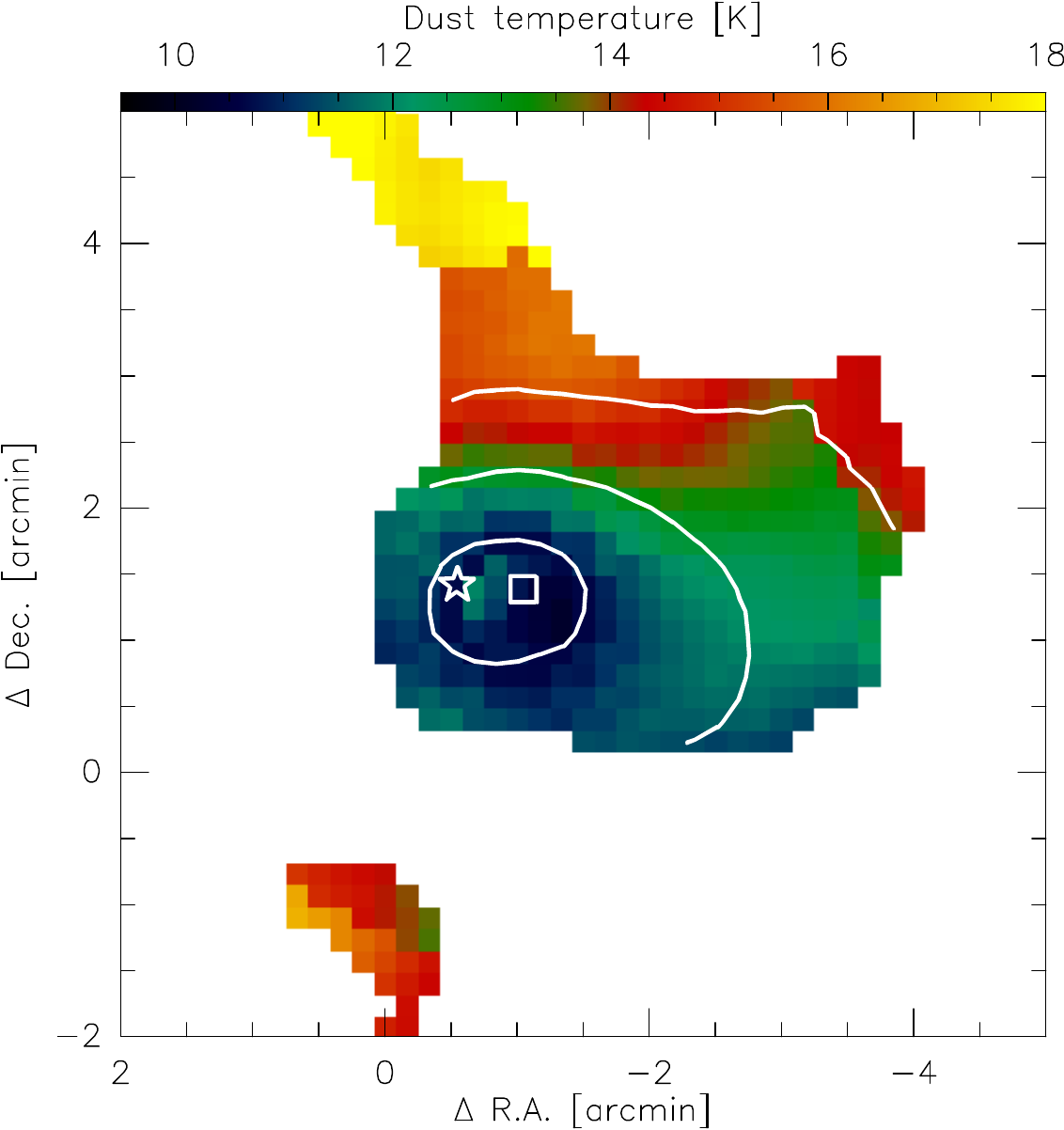}
   \caption{\footnotesize{Dust temperature map of CB130 overlaid with contours of the hydrogen density. They mark densities of $10^3$, $10^4$, and $10^5$~cm$^{-3}$. The square indicates the center of the
     starless core.}}
   \label{fig_dustmaps_CB130}
\end{figure}

\begin{figure}[htpb]
   \centering
   \includegraphics[width=0.45\textwidth]{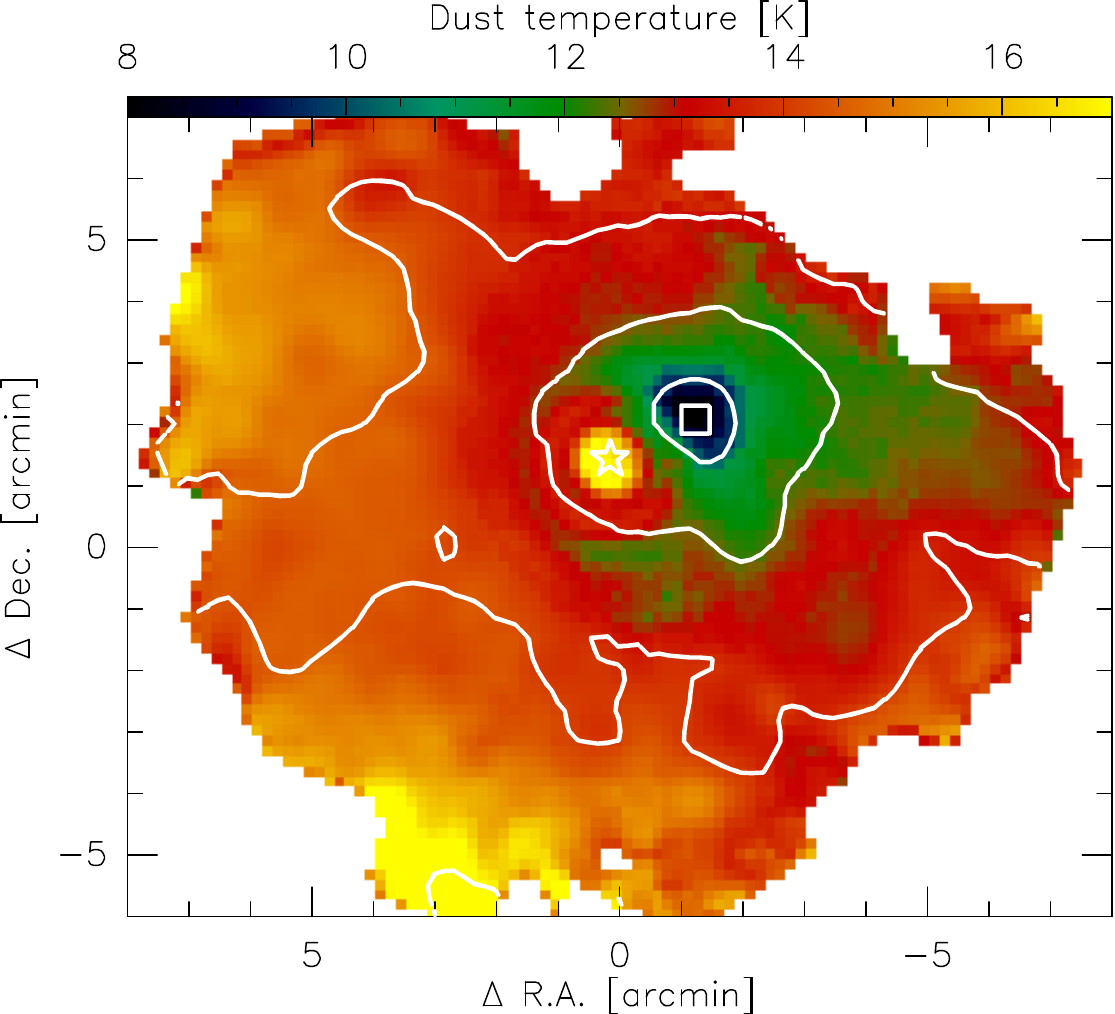}
  \caption{\footnotesize{Dust temperature map of CB244 overlaid with contours of the hydrogen density. They mark densities of $10^3$, $10^4$, and $10^5$~cm$^{-3}$. The square indicates the center of the
     starless core.}}
   \label{fig_dustmaps_CB244}
\end{figure}

}
..

\pagebreak[4]
..
\clearpage
\section{LTE Analysis}
\label{sec:LTE}

\subsection{Model}

We estimated the molecular column densities of $^{13}$CO, C$^{18}$O, and $\mathrm{N_2H^+}$ in a first step with a simple and
well-established approach. For completeness, we briefy describe here this simplified approach and  the results and discuss the differences and drawbacks with respect to the full chemical modeling.

We do not model the emission of the $^{12}$CO~(J=2-1) transition, since it becomes optically thick already at $n_\mathrm H < 10^4$ and therefore does not trace the regions of expected freezeout. At these low densities, gas and dust temperatures might also be decoupled \citep{galli2002}, which is an additional obstacle for interpreting the $^{12}$CO emission.
[p]
The method is described in \citet{stahler05}. It assumes local thermal equilibrium~(LTE)
and a constant gas temperature along the line-of-sight. Since we do not have an
independent measurement of the gas temperature, we make the simplifying
assumption that the kinetic gas temperature is the same as the dust temperature we obtained from the LoS-averaged black-body-fitting of the continuum data
in \citet{launhardt2013}. This is justified in the dense interiors of the
starless cores, but may no longer be strictly valid in the outer parts at
hydrogen densities below a few $10^4$~cm$^{-3}$.

To prepare the analysis, we obtain maps of the flux in the observed lines by integrating over the velocity axis of the
spectral cubes. Maps of the linewidths of the CO isotopologues were derived
from Gaussian fits to the spectra. For $\mathrm{N_2H^+}$ we use the
hfs-fitting routine provided with the
\textit{CLASS} package of \textit{GILDAS}\footnote{http://www.iram.fr/IRAMFR/GILDAS}. This routine allows fitting the full hyperfine structure of the $J=1-0$ transition and thus also
derives the optical depth of the lines. The frequency offsets and relative
strengths of the individual lines are adopted from \citet{caselli1995}. With
these intermediate results at hand we can derive the molecular column
densities using
\begin{equation} 
  N=\frac{8\pi\nu_0^2\Delta \nu Q \Delta \tau}{c^2[p]
    A_\mathrm{ul}}\left(\frac{g_\mathrm l}{g_\mathrm u}\right)\left[1-\exp\left(-\frac{T_0}{T_{\mathrm{ex}}}\right)\right]^{-1},
\end{equation}
where $\nu_0$ is the frequency of the transition, $\Delta \nu$ the observed
linewidth, $Q$ the partition function of the rotational levels, $\Delta \tau$
the total optical thickness of the line, $A_{ud}$ the Einstein parameter of
the transition, $g_i$ the relative statistical weights of the upper and lower
levels, $T_0=h\nu_0/k_B$, and $T_\mathrm{ex}$ the excitation temperature of
the molecules. Since we assume LTE, we set
$T_\mathrm{ex}=T_\mathrm{gas}=T_\mathrm{dust,\,MBB}$. The partition function
$Q$ is given by

\begin{equation}
Q=\sum_J \left(2J+1\right)\exp\left[-\frac{h}{k_\mathrm BT_\mathrm{ex}}\left(BJ\left(J+1\right)\right)\right],
\end{equation}
where $B=55101.011$~MHz is the rotational constant for $^{13}$CO,
$B=54891.420$~MHz for C$^{18}$O, and $B=46586.867$~MHz for
N$_2$H$^+$~\citep{pickett1998}. While the optical depth of $\mathrm{N_2H^+}$
could be determined directly by comparing the strength of the individual
hyperfine components, we need to apply the ``detection equation'' in order to
derive the optical depth of the CO lines:

\begin{equation}
T_B = T_0\left[f(T_\mathrm{ex})-f(T_\mathrm{bg})\right]\left[1-\mathrm e^{-\Delta
\tau} \right],
\end{equation} 
where $f(T)\equiv\left(\exp(T_0/T)-1\right)^{-1}$,
$T_\mathrm{bg}=2.73$~K, and $T_\mathrm B$~taken as the peak value of the Gaussian fits.

One-dimensional profiles of the resulting maps were obtained
by azimuthally averaging around the core centers. In case of
asymmetries, a segment of the circle in this direction has been masked
for the averaging task. The regions that have been taken into account
for deriving the 1D-profiles are indicated in
Figs.~\ref{fig_obs_cb4}~-~\ref{fig_obs_cb244}.
\subsection{Results}

\begin{figure*}
   \centering
   \includegraphics[width=0.85\textwidth]{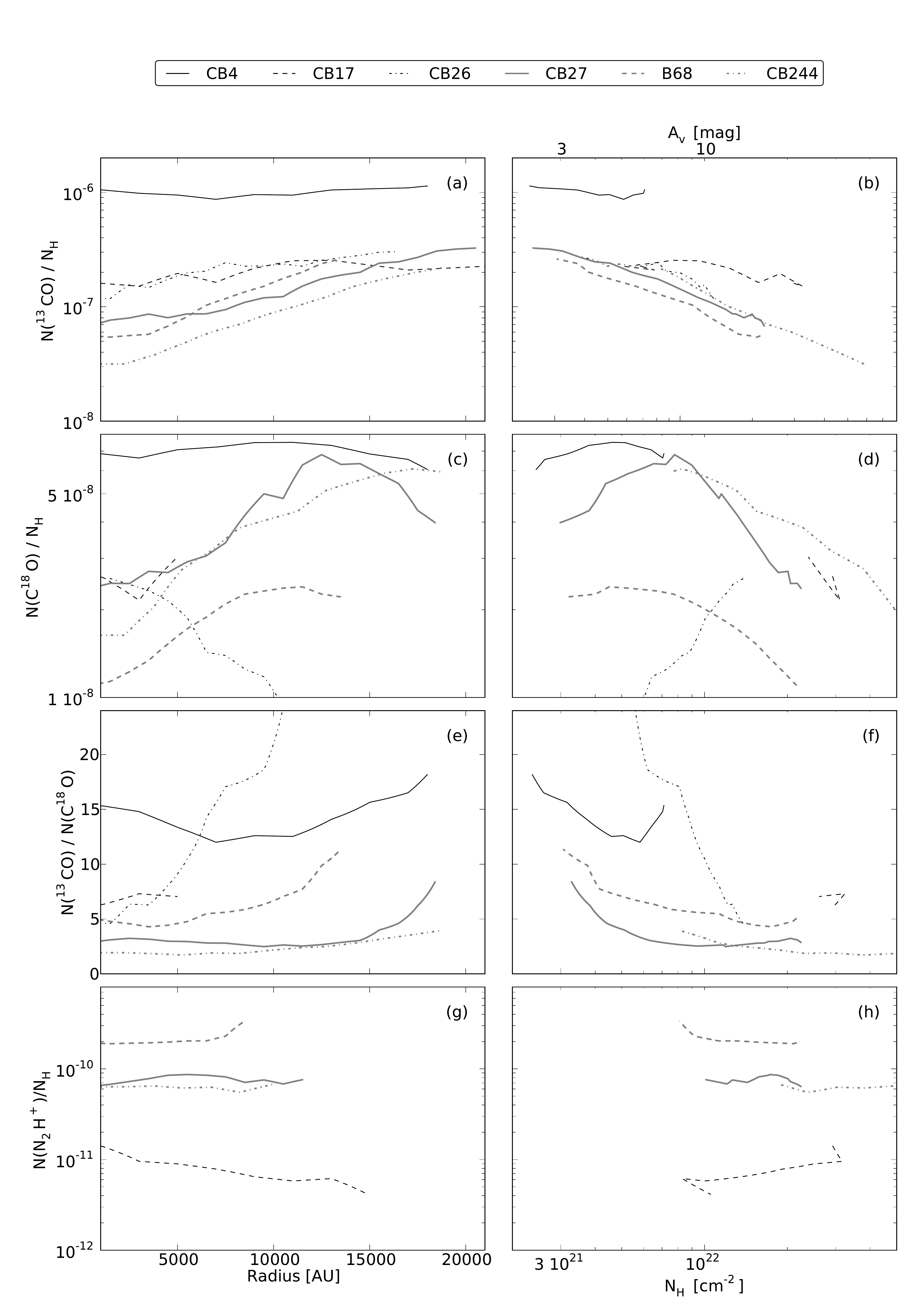}
   \caption{\footnotesize{Results from the LTE-analysis. Plotted are \textit{(a)} the relative column densities of $^{13}$CO against radius \textit{(b)} and against the hydrogen column density, \textit{(c)} the relative column densities of C$^{18}$O against radius \textit{(d)} and against the hydrogen column density, \textit{(e)} the ratios of N($^{13}$CO)/N(C$^{18}$O) are plotted against radius \textit{(f)} and against the hydrogen column density, \textit{(g)} the relative column densities of N$_2$H$^+$against radius,  \textit{(h)} and against the hydrogen column density. The column densities of N$_2$H$^+$ in CB\,4 and CB\,26 could not be derived, since the emission in these globules is too weak so are missing in the plots.}}
   \label{fig_LTE_1D}
\end{figure*}
The derived radial profiles of the molecular column densities of $^{13}$CO, C$^{18}$O ,and N$_2$H$^+$ are presented in Figure~\ref{fig_LTE_1D}. We find signs of depletion of $^{13}$CO in all cores except for CB\,4, which is the most tenuous core of the studied sample. The $^{13}$CO column density of the other cores decreases constantly with the increasing hydrogen column density. The results for $^{13}$CO are, however, affected by optical depth effects, as well as deviations of
the gas temperature from the LoS-averaged dust temperature and sub-thermal rotational excitation.

The derivation of C$^{18}$O column densities is less influenced by
these effects since it is a rarer isotopologue (a factor of $\sim 7$ lower abundance as compared to $^{13}$CO) and since its emission comes from more restricted regions
with higher gas densities. The plots of the $^{13}$CO column densities vs. visual extinction~(top left panel of Fig.~\ref{fig_LTE_1D}) yield a remarkable difference to those of the C$^{18}$O column densities. The relative abundance of $^{13}$CO drops continuously with increasing
extinction, indicating central depletion of this species. While this behavior
is also found for C$^{18}$O, its relative abundance also drops toward low
hydrogen column densities and peaks between an $A_v$ of 5~to~10~mag. The drop
toward lower extinction can be explained by a weaker self-shielding of this
rare isotopologue, which in turn leads to photodissociation of C$^{18}$O in the envelopes where $^{13}$CO is already well shielded. The chemical modeling, however, suggests that the UV radiation is already attenuated by several magnitudes in these regions compared to the nominal strength of the ISRF~(see Sect.~\ref{sec:uv}). The difference in the profiles of both species becomes more obvious from the ratio of both column densities. The ratio
$N(\mathrm{^{13}CO})/N(\mathrm{C^{18}O})$ decreases continuously with increasing
hydrogen column density in all starless cores. Interestingly, it even drops below
the usually assumed ratio of 7 for the interstellar medium~(ISM). The change
in the $N(\mathrm{^{13}CO})/N(\mathrm{C^{18}O})$ ratio could thus not only be
due to different self-shielding of both species but also partly due to
reactions that increase the abundance of C$^{18}$O with respect to $^{13}$CO
due to ion-molecule exchange reactions in cold and dense environments \citep[e.g.,][]{langer1984}.

Where the $\mathrm{N_2H^+}$ emission is strong enough to fit the hyperfine structure, we also derive column densities for this molecule. We find that the ratio N($\mathrm{N_2H^+}$)/N$_\mathrm{H}$ is roughly constant within the globules~(see bottom row of Fig.~\ref{fig_LTE_1D}). This finding contrasts with the results of our advanced approach using chemical modeling and a subsequent line-radiative transfer. There we find depletion of $\mathrm{N_2H^+}$ in the centers of the majority of the globules~(see Sect.~\ref{sec:chem_n2hp} and Fig.~\ref{fig_abundances}). This demonstrates the limitation of the often-used LOS-averaged analysis.



\end{appendix}

\end{document}